\documentclass[12pt]{article}
\usepackage[utf8]{inputenc}
\usepackage{graphicx}
\usepackage[export]{adjustbox}
\usepackage{wrapfig}
\usepackage{subfigure}
\usepackage{caption}
\usepackage{geometry}
\usepackage{amsfonts}
\usepackage{physics}
\usepackage{float}
\usepackage{dynkin-diagrams}
\usepackage{amsmath,amssymb,amsthm,amscd}
\usepackage[hidelinks]{hyperref}

\usepackage{csvsimple}
\usepackage{comment}
\usepackage{afterpage}

\usepackage{slashed}                      
\usepackage{tikz}   

\usepackage{blindtext}
\usepackage{wrapfig}

\usepackage{url}  
\usepackage{longtable}
\usepackage{rotating} 
\usepackage{multirow}
\usepackage{ulem}
\usepackage{color}

\newcommand{\rd}{{\mathrm{d}}}

\def \be {\begin{equation}}
\def \ee {\end{equation}}
\def \ba {\begin{aligned}}
\def \ea {\end{aligned}}

\begin{document}
\begin{titlepage}
\begin{flushright}
			TIT/HEP-695 \\
			July 2023
		\end{flushright}
		\vspace{0.1cm}
  \begin{center}
			{
   \bf
				Exact WKB Analysis and TBA Equations for the Stark Effect
			}
			\vskip 1cm
			{\large  Katsushi Ito\footnote{ito@th.phys.titech.ac.jp} and Jingjing Yang\footnote{j.yang@th.phys.titech.ac.jp} }
			\vskip 1.0em
			{\normalsize\it Department of Physics,\\
				Tokyo Institute of Technology\\
				Tokyo, 152-8551, Japan}
			\vskip 1.0em
		\end{center}
		\begin{abstract}
We apply the exact WKB analysis to a couple of one-dimensional Schr\"{o}dinger-type equations reduced from the Stark effect of hydrogen in a uniform electric field. By introducing Langer's modification and incorporating the Stokes graphs, we prove the exactness of the Bohr-Sommerfeld quantization conditions for the Borel-resummed quantum WKB periods in the specific parameter regions of the electric field intensity and magnetic quantum number. It is also found these quantization conditions get modified with an additional suppressed contribution when the parameters vary beyond the specific regions. We also present Thermodynamic Bethe Ansatz (TBA) equations governing the quantum periods in the absence of Langer's modification and discuss its wall-crossing and analytic continuation. Numerical calculations are conducted to compare the complex resonant frequencies from our quantization conditions against ones from the Riccati-Padé method, the TBA equations are also confirmed by comparing its expansions with all-order quantum periods.
\end{abstract}
\end{titlepage}

\tableofcontents
\newpage
\baselineskip=0.7cm
 	\numberwithin{equation}{section}
	\numberwithin{figure}{section}
	\numberwithin{table}{section}
\section{Introduction}
The Stark effect describing hydrogen under an external electric field \cite{stark}, is one of the earliest examples considered since the invention of quantum mechanics. The external field leads bound states for hydrogen to resonant states, whose energy becomes complex $E=E_c+i\Gamma/2$. $E_c$ represents the peak of the energy spectra and $\Gamma$ stands for the width related to the ionization rate. There are many approaches based on perturbation theory to solve this problem \cite{Schrodinger, Epstein26,pert78, Damburg1976, Kolosov83, Dolgov}. But the perturbative expansions for the Stark effect are asymptotic, so additional numerical methods such as the Borel-Pad{\'e} technique are necessary. There are also WKB-type approaches that rely on the fact that the three-dimensional Schr\"{o}dinger equation for the Stark effect is separable under parabolic coordinates. Both of the separated equations take the form of the one-dimensional Schr\"{o}dinger equation with a linear and inverse potential in addition to a centrifugal term. To satisfy the correct boundary condition at origin, Kramers and especially Langer \cite{Langer37} introduce the so-called Kramers-Langer substitution or Langer's modification used in this paper, see \cite{langer3}. A comprehensive interpretation can be found in \cite{langer2} in terms of two-$\hbar$ notation in an enlightening perspective. The successive research on the WKB method for the Stark problem is almost based on Langer's modification, such as \cite{starkwkb62,starkwkb69,starkwkb82,starkwkb99}. These works mainly rely on the period integrals in the classically allowed region, which will be notated as classical WKB periods.

The main purpose of this paper is to study the Stark effect by the exact WKB analysis. This can be achieved by the Borel resummation of formal $\hbar$ power series in the framework of resurgence theory \cite{Ecalle, BPV, Voros, Voros1}. The analytic behaviour of the Borel-resummed WKB solution is determined by Stokes graphs, and the relation of WKB solutions for different Stokes regions is described by connection formulas \cite{kawai,cluster1,cluster2}. In quantum mechanical problems, one imposes boundary conditions for the wave functions, these deduce the exact quantization conditions (EQCs) for the eigenvalue problems in terms of quantum period \cite{Silverstone81, silverstone, sueishi1,sueishi2}. It is Borel-resummed quantum WKB periods appearing to connect the solutions with different normalizations. 
Its discontinuity is
captured by the Delabaere-Pham formula (DP formula) \cite{DDP22, DP}. 
In this paper, we apply the exact WKB analysis to the Stark effect to obtain the exact quantization conditions.

The development of the WKB analysis invoked the study of integrable models and is formed into the so-called ODE/IM correspondence \cite{DT, DT2}. The ODE/IM correspondence indicates there is a relationship between the $Y$-functions in integrable models and quantum periods from the WKB analysis. Specifically, $Y$-systems derived from the Wronskian relations coincide with equations followed by quantum periods from the DP formula \cite{IMS}. These studies introduce new techniques into integrable models and are fed back to the research on the ODE side. Especially, the ODE/IM correspondence introduces TBA equations for quantum periods, which provides a non-perturbative completion for these asymptotically defined quantities in the WKB analysis. This correspondence was summarized explicitly for polynomial potentials \cite{IMS, Emery} and also potentials with regular singularity \cite{Ito-Shu}. As one analytically continues the parameters in the potential, TBA equations get modified when wall-crossing of quantum periods happens \cite{GMN}. 
We would like to consider these kinds of TBA equations corresponding to the reduced one-dimensional equations for the Stark effect.

This paper is organized as follows. In section \ref{sec:wkb}, we review the fundamentals of the exact WKB analysis. Our work starts from section \ref{sc: langer}, which focuses on the application of the exact WKB analysis to the Stark effect.
We consider Langer's modification and all-order quantum corrections and write down the exact quantization conditions (EQCs) from the Stokes graph. It is found the Bohr-Sommerfeld quantization condition for the Borel-resummed periods is already exact in the specific parameter regions but gets an additional term when the parameters vary beyond these regions. Our EQCs are numerically tested against the Riccati-Pad\'e method \cite{rpm}. Section \ref{sc:tba} is devoted to finding integral TBA equations for quantum periods. However, these TBA equations are established for bare effective potentials without Langer's modification, which is different from the prescription adopted in section \ref{sc: langer}. We discussed the wall-crossing of the TBA equations \cite{GMN} and also the analytic continuation \cite{DT3,xiyin} against the parameters for the Stark effect. We conclude our results in section \ref{sc:concl}.

\section{Exact WKB analysis and connection formulas}
\label{sec:wkb}
In this section, we give a brief review to mention fundamental aspects and fix the notations of the exact WKB analysis. We intensively  consult the work of \cite{kawai,cluster1,cluster2,sueishi1,sueishi2} and the references therein for this section.

\subsection{Review of the all-order WKB expansions}
The Stark effect involves a couple of one-dimensional Schr{\"o}dinger equations which will be discussed in section \ref{sc: langer}. Let us consider the equation for a generic potential $Q(x,\hbar)$ 
\begin{equation}
\label{eq:seq}
\left(-\hbar^{2} \frac{d^{2}}{d x^{2}}+Q(x,\hbar) \right) \psi(x)=0,
\end{equation}
where 
\begin{equation}
Q(x,\hbar)=Q_0(x)+Q_2(x)\hbar^2
\end{equation}
consists of one classical potential $Q_0(x)$, and an additional $\hbar$-dependent term $Q_2(x)\hbar^2$, usually called quantum potential. In the exact WKB analysis, the coordinate $x$ and the reduced Planck constant $\hbar$ are taken to be complex variables, and $\hbar$ is regarded as an expansion parameter. One starts the WKB analysis by setting an ansatz
\begin{equation}
\label{eq:ansatz}
    \psi(x)=\exp\left( \frac{1}{\hbar}\int^{x}P(x,\hbar)dx\right),
\end{equation}
where 
\begin{equation}
    P(x,\hbar)=\sum_0^\infty p_n(x)\hbar^n
\end{equation}
is a formal power series. Substituting the above solution for wave function into \eqref{eq:seq}, then we arrive at the Riccati equation
\begin{equation}
\label{eq:riccati}
    \hbar P^{\prime}(x)=Q_0(x)-P^2(x)+\hbar^2Q_2(x).
\end{equation}
This equality holds to all orders, matching order by order reduces to the following recursive relations for $p_n(x)$
\begin{equation}
\label{eq: recursion}
    p_n(x)=-\frac{1}{2p_0(x)}\left(\sum_{i=1}^{n-1}p_{n-i}(x)p_i(x)+p_{n-1}^{\prime}(x)\right), \quad n\geq3,
\end{equation}
with 
$$
p_0^2(x)=Q_0(x),\quad
p_1(x)=-\frac{p_0^{\prime}(x)}{2p_0(x)}, \quad p_2(x)=-\frac{1}{2p_0(x)}\left(p_1^{2}(x)+p_1^{\prime}(x)-Q_2(x)\right).
$$
The odd and even power of $P(x,\hbar)$
\begin{equation}
     P_{\mathrm{even}}(x,\hbar)=\sum_{0}^{\infty}p_{2n}(x)\hbar^{2n},\quad
      P_{\mathrm{odd}}(x,\hbar)=\sum_{0}^{\infty}p_{2n+1}(x)\hbar^{2n+1},
\end{equation}
was related by
\begin{equation}
    P_{\mathrm{odd}}(x,\hbar)=-\frac{\hbar}{2}\frac{d}{d x}\log P_{\mathrm{even}}(x,\hbar).
\end{equation}
Hereafter we omit $\hbar$-dependence in function $P(x,\hbar)$. Then the WKB solution \eqref{eq:ansatz} can be expressed in terms of $P_{\mathrm{even}}(x)$ as
\begin{equation}
\label{eq:wkb}
    	\psi_{a}^{\pm}(x)=\frac{1}{\sqrt{P_{\mathrm{even}}(x)}}\exp\left( \pm\frac{1}{\hbar}\int_{a}^{x}P_{\mathrm{even}}(x)d x\right),
\end{equation}
where $a$ is a reference point for the integral, and is often exclusively chosen as a turning point, that satisfies $Q_{0}(a)=0$. The prescription $\pm$ in the exponential signifies the properties of the solutions, exponentially suppressed, exponentially growing, or oscillatory in the given region of $x$. The generic solution normalized at $a$ consists of the linear combination of two independent components $\psi_{a}^{\pm}(x)$
\begin{equation}
    	\psi_{a}(x)=\frac{C_{+}}{\sqrt{P_{\mathrm{even}}(x)}}\exp\left( \frac{1}{\hbar}\int_{a}^{x}P_{\mathrm{even}}(x) d x\right)+
    	\frac{C_{-}}{\sqrt{P_{\mathrm{even}}(x)}}\exp\left( -\frac{1}{\hbar}\int_{a}^{x}P_{\mathrm{even}}(x) d x\right).
\end{equation}
One can formally include the all-order quantum corrections into the wave function by expanding \eqref{eq:wkb} to orders of $\hbar$
\begin{equation}
\label{eq:wkbexpansion}
\psi_{a}^{\pm}(x)=\exp\left(\pm \frac{1}{\hbar} \int_{a}^{x} p_0(x) d x\right)\sum_{n=0}^{\infty} \psi_{a, n}^{\pm}(x) \hbar^{n}.
\end{equation}
Hereafter, we specify $p_0(x)=\sqrt{Q_0(x)}$. This formal series includes an exponential prefactor and formal power series with its coefficients $x$-dependent. It is of course divergent in general. The Borel resummation provides a procedure to deal with this sort of divergent series. We will consider its utilization into formal series \eqref{eq:wkbexpansion}.

\subsection{Stokes graphs and connection formulas}

\subsubsection{Borel resummation and Stokes graphs}
\label{sc:resum}

\begin{figure}[H]
        \begin{minipage}{0.48\textwidth}
         \centering
\includegraphics[width=1\linewidth]{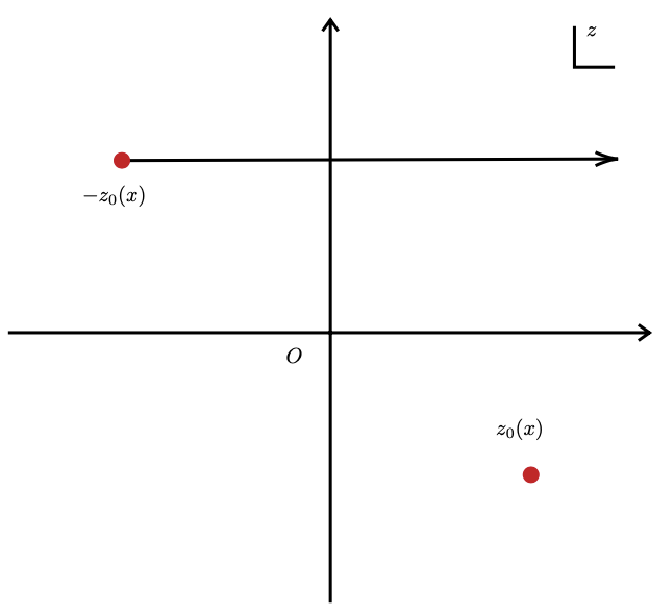}
    \caption{This is the Borel plane for Borel transform of $\psi_a^{+}(x)$. The arrow represents the integral contour for $\theta=0$. The right red point denotes the corresponding Borel pole.}
     \label{fig:wkbborel}
    \end{minipage}
    \hspace{3mm}
        \begin{minipage}{0.48\textwidth}
         \centering
\includegraphics[width=1\linewidth]{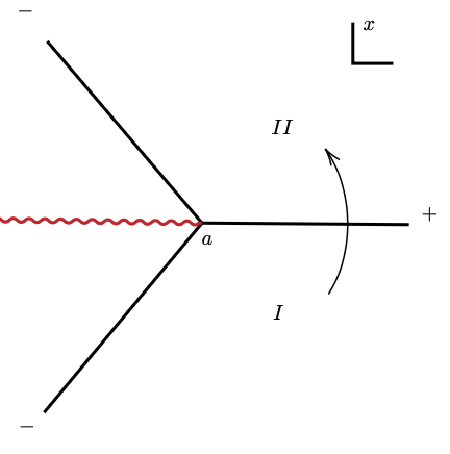}
     \caption{This figure shows the Stokes lines emanating from a simple turning point $a$. $\pm$ labels the orientations of the Stokes lines. The arrow indicates the continuation path from Stokes region $\mathrm{I}$ to $\mathrm{II}$ and the red wavy line is the branch cut.}
      \label{fig: airy}
    \end{minipage}
\end{figure}

We consider a generic form of formal series with exponential prefactor, which appears at the formal expansion of the WKB solution for wave function \eqref{eq:wkbexpansion}
\be
\label{eq:formalseries}
f(\hbar)=e^{-\frac{A}{\hbar}}\sum_{n=0}^{\infty} a_{n} \hbar^{n+\alpha}, \quad \alpha \notin \{-1,-2,-3,\cdots\},
\ee
where $A$ is a constant independent of $\hbar$ and $a_n$ diverges factorially $a_n\sim n!$. Its Borel transform is  
\be
\label{eq:boreltransf}
    \mathcal{B}[f](z)=\sum_{n=0}^{\infty}\frac{a_n}{\Gamma(n+\alpha)}(z-A)^{n+\alpha-1},
\ee
which transforms the divergent formal series \eqref{eq:formalseries} into a well-defined function in the new variable $z$. This function has a finite radius of convergence and can be analytically continued to a domain including the real $z$-axis, with possible singularities, such as poles. We often call this complex $z$-plane the Borel plane. If there is no singularity along the positive real line, one can define the Laplace transform of Borel-transformed function $\mathcal{B}[f](z)$ as
\be
    \mathcal{S}[f](\hbar)=\mathcal{L}[\mathcal{B}[f](z)]=\int_{A}^{\infty} dz e^{-\frac{z}{\hbar}} \mathcal{B}[f](z),
\ee
it is the Borel resummation or simply Borel resum of the original formal series \eqref{eq:formalseries}, and we say the formal series \eqref{eq:formalseries} is Borel summable in this case. 
The directional Borel resummation can be defined similarly if there is no singularity in the direction with angle $\theta$
\be
\label{eq:borelsum}
    \mathcal{S}_{\theta}[f](\hbar)=\mathcal{L}_{\theta}[\mathcal{B}[f](z)]=\int_{A}^{e^{i \theta}\infty} dz e^{-\frac{z}{\hbar}} \mathcal{B}[f](z).
\ee
One may note this sort of directional summation is equivalent to writing as 
\begin{equation}
     \mathcal{S}_{\theta}[f](\hbar)=\mathcal{S}_0[f](e^{i \theta}\hbar),
\end{equation}
which indicates the Borel resummation for formal series at the angle $\theta$ is equivalent to the resummation at angle $0$ by rotating $\hbar$ to $e^{i\theta}\hbar$. We will omit the subscript $0$ in $\mathcal{S}_{\theta}$ when $\theta=0$. It is much more fruitful to look at the singularities in the Borel plane, let us simply consider there is a single pole along $\theta$ ray, then the integral defined in \eqref{eq:borelsum} makes no sense. But we can bypass the pole by small deformation from above and below, this suggests defining the lateral Borel resummation for \eqref{eq:formalseries} as 
\be
\label{eq:phi3}
    \mathcal{S}_{\theta\pm}[f](\hbar)=\lim_{\delta \to 0}\mathcal{L}_{\theta\pm \delta}[\mathcal{B}[f](z)].
\ee
Let us now consider the formal series expansion for wave function \eqref{eq:wkbexpansion}. We will study its Borel singularities and Borel summability. The general Borel resummation procedure is applicable, but the expansion coefficients $\psi_{a,n}^{\pm}$ are parameter-dependent on $x$. This parameter-dependence promotes the singularities in the Borel plane to be $x$-dependent and then defines curves in coordinate $x$ plane, which are Stokes lines introduced later. At first, we do Borel transform of  \eqref{eq:wkbexpansion} to suppress the divergence
\begin{equation}
\label{eq:borelpsi}
\mathcal{B}\left[\psi_{a}^{\pm}(x)\right](z)=\sum_{n=0}^{\infty} \frac{\psi_{a, n}^{\pm}(x)}{\Gamma(n)}\left(z \pm z_{0}\right)^{n-1}, \quad z_{0}=\frac{1}{\hbar}\int_{a}^{x} p_0(x) \rd x.
\end{equation}
Now we obtain a convergent function defined in some domain in the $z$-plane, it is generally meromorphic with possible singularities and one can further continue the function into the whole $z$-plane. Its Laplace transform can be defined accordingly unless the integral contour encounters Borel poles.
\begin{equation}
\mathcal{S}_{\theta}\left[\psi_{a}^{\pm}\right](\hbar)=\int_{\mp z_{0}}^{e^{i\theta}\infty } e^{-\frac{z}{\hbar}} \mathcal{B}\left[\psi_{a}^{\pm}(x)\right](z) d z, \quad \theta=\arg(\hbar).
\end{equation}
Figure \ref{fig:wkbborel} shows the definition of $\mathcal{S}_{\theta}[\psi_a^{+}](\hbar)$ for $\theta=0$. Hereafter we will use $\psi_{a}^{\pm}(x)$ to represent Bore-resummed wave functions for simplicity without confusion. It is proved that $\pm z_0(x)$ are singular points of the Borel transform \eqref{eq:borelpsi} \cite{kawai}. The location of $\pm z_{0}(x)$ depends on the coordinate $x$, so the integral contour can touch the Borel pole $z_0(x)$ in figure \ref{fig:wkbborel} by moving in the complex $x$ plane. This condition can be expressed as
\begin{equation}
\label{eq:stokescurve}
\operatorname{Im} \frac{1}{\hbar} \int_{a}^{x} p_0(x) \rd x=0,
\end{equation}
which defines Stokes lines emanating from a turning point $a$. Moreover, the real part tells us which one in $\psi_{a}^{\pm}(x)$ is growing or decreasing, so we attach each line $+$ or $-$ to specify the behaviour of solutions. It is determined by the sign of the following quantity
\begin{equation}
\operatorname{Re} \frac{1}{\hbar} \int_{a}^{x} p_0(x) \rd x.
\end{equation}
The positive value indicates $\psi^{+}(x)$ increases exponentially, while $\psi^{-}(x)$ decreases exponentially. If it is negative, $\psi^{-}(x)$ increases while $\psi^{+}(x)$ decreases. 
For each turning point, one can define the corresponding Stokes lines, and there are three Stokes lines for a simple turning point, i.e. the simple zero of $Q_0(x)$. Figure \ref{fig: airy} shows the Stokes lines emanating from simple turning point $a$ to complex infinity. One Stokes line might start from one turning point, the origin or infinity, and end at another turning point, the origin or infinity. The Stokes line connecting two turning points is degenerate and can be resolved by saddle reduction via introducing a small deformation for integral contour $\theta\to \theta\pm\delta$.

The Stokes lines divide the coordinate plane into several adjacent regions, called Stokes regions. The collection of all the Stokes lines composes Stokes graphs. In each Stokes region, the formal series of the WKB solutions is Borel summable to define the analytic solution for each region. However, if one continues the solution in one region to another adjacent region by crossing a Stokes line, one of the components in $\psi^{\pm}$ gets a discontinuity while the other one does not. This is quantitatively described by the connection formula in the next subsection. 

\subsubsection{Connection formulas and quantum periods}
As discussed in section \ref{sc:resum}, the Borel singularities for WKB solutions are captured by the Stokes lines. These singularities introduce discontinuities when the WKB solution in one Stokes region is continued across the Stokes line to an adjacent region. In general, this property is concluded by the following connection formula \cite{cluster2,sueishi1}.
\begin{itemize}
    \item If crossing a Stokes line anticlockwise at which the $\psi^+$ is dominant, $\psi^+$ gains a discontinuity controlled by $\psi^-$, while $\psi^-$ doesn't change,
    \be
    \label{eq:connection1}
 \psi^+\to \psi^++i\psi^-,\quad  \psi^-\to \psi^-,
    \ee    
    As aforementioned, the symbol for the Borel resum is omitted.     
    \item If crossing the Stokes line anticlockwise at which the $\psi^-$ is dominant, $\psi^-$ gains a discontinuity controlled by $\psi^+$, while $\psi^+$ remains unchanged,
    \be
    \label{eq:connection2}
   \psi^-\to \psi^-+i\psi^+,\quad  \psi^+\to \psi^+,
    \ee    
    \item If crossing a Stokes line clockwise, the formulas differ by flipping the $+$ to $-$ in front of the $i$, 
     \item If crossing a branch cut emerging from a turning point anti-clockwise (clockwise), the dominant and subdominant component exchange (with a minus sign),
      \be
 \psi^+\to \pm \psi^-,\quad  \psi^-\to \pm \psi^+,\, +\, \text{for anticlockwise},\, -\, \text{for clockwise}.
    \ee 
\end{itemize}
One defines the matrix $M_{+}$ and $M_{-}$ to represent \eqref{eq:connection1} and \eqref{eq:connection2}
\be
M_{+}=
\left(\begin{array}{ll}
1 & i \\
0 & 1
\end{array}\right),\quad
M_{-}=
\left(\begin{array}{ll}
1 & 0 \\
i & 1
\end{array}\right),
\ee
which is called the monodromy matrix. Then the connection formulas state that the WKB solutions in one Stokes region can be obtained via multiplication of the monodromy matrix by the solutions in the adjacent region, specifically
\begin{equation}
\left(\begin{array}{c}
\psi_{a, \mathrm{I}}^{+} \\
\psi_{a, \mathrm{I}}^{-}
\end{array}\right)=M\left(\begin{array}{l}
\psi_{a, \mathrm{II}}^{+} \\
\psi_{a, \mathrm{II}}^{-}
\end{array}\right),
\end{equation}
which is represented in figure \ref{fig: airy} where $M=M_+$.

Furthermore, the two independent wave functions normalized in a turning point $a_1$ denoted as $\psi_{a_1}^{\pm}$, and $\psi_{a_2}^{\pm}$ normalized in another turning point $a_2$ are related by normalization coefficients $N_{a_{1}a_{2}}^{\pm}$ defined as
\begin{equation}
N_{a_{1}a_{2}}^{\pm}=\exp\left( \pm\frac{1}{\hbar}\int_{a_1}^{a_2}P_{\mathrm{even}}(x)d x\right),
\end{equation}
which induces the following connection formula to relate wave functions normalized in $a_2$ to $a_1$
\begin{equation}
\label{eq:tconnection}
    \psi_{a_1}^{\pm}(x)=N_{a_{1}a_{2}}^{\pm} \psi_{a_2}^{\pm}(x).
\end{equation}
In the latter context, we will intensively use the following notation
\be
\label{eq:vorossymbol}
\mathcal{V}_{\gamma}\equiv \exp\left(\frac{1}{\hbar}\oint_{\gamma} P_{\mathrm{even}}(x)d x\right),
\ee
where $\gamma$ is the cycle encircling two turning points $a_1$ and $a_2$. It is usually called Voros multipliers or Voros symbols in the literature. As the same as  the monodromy matrix, we can express the above connection formula in the matrix form, which transforms the wave function basis at turning point $a_2$ to $a_1$
\begin{equation}
\left(\begin{array}{c}
\psi_{a_{1}}^{+}(x) \\
\psi_{a_{1}}^{-}(x)
\end{array}\right)=N_{a_{1} a_{2}}\left(\begin{array}{l}
\psi_{a_{2}}^{+}(x) \\
\psi_{a_{2}}^{-}(x)
\end{array}\right),\quad 
N_{a_{1} a_{2}}=\left(\begin{array}{cc}
\mathcal{V}_{\gamma}^{\frac{1}{2}} & 0 \\
0 & \mathcal{V}_{\gamma}^{-\frac{1}{2}}
\end{array}\right).
\end{equation}
At the equal significance, we rename the quantity $\oint_{\gamma} P_{\mathrm{even}}(x)d x$ the quantum period or quantum WKB period corresponding to the cycle $\gamma$
\begin{equation}
\label{eq:qp}
    \Pi_{\gamma}(\hbar)=\oint_{\gamma}P_{\mathrm{even}}(x)d x.
\end{equation}
It is now necessary to re-explain notations. Corresponding to~\eqref{eq:seq}. There exists a classical curve 
\begin{equation}
    y^{2}=Q_0(x),
\end{equation}
it is called the WKB curve. Geometrically it defines a Riemann surface $\Sigma_{\mathrm{WKB}}$, and the quantum period is the integral of meromorphic differential $P_{\mathrm{even}}(x) d x$ along the one-cycle $
\gamma \in H_{1}\left(\Sigma_{\mathrm{WKB}}\right)
$. The leading order of the quantum period is often called the classical WKB period or classical period
\begin{equation}
     \Pi_{\gamma}^{(0)}\equiv \oint_{\gamma}p_0(x)d x,
\end{equation}
which plays an important role in the history of quantum mechanics. The well-known Bohr-Sommerfeld (BS) quantization condition can be written in terms of the classical period as
\begin{equation}
\label{eq:hoeqc}
    \Pi_{\gamma}^{(0)}=2\pi i \hbar \left(n+\frac{1}{2}\right),\, n=0,1,2,\cdots,
\end{equation}
where $\gamma$ is specified to be one cycle around the classically allowed region. It is widely used for two-turning-point problems by accounting for the classical period integral encircling turning points $a_1$ and $a_2$ in the classically allowed region. Here $i$ appears since our notation for $p_0(x)=\sqrt{V(x)-E}$, which is different from the conventional choice $p_0(x)=\sqrt{E-V(x)}$.

We note that $P_{\mathrm{even}}(x)$ is defined by recursive relation \eqref{eq: recursion}. Then the Voros symbols \eqref{eq:vorossymbol} and quantum periods \eqref{eq:qp} are asymptotic series in $\hbar$
\begin{equation}
    \Pi_{\gamma}(\hbar)= \oint_{\gamma} \left(\sum_{n=0}^{\infty} p_{2n}(x,\hbar) \hbar^{2n}\right) d x\sim\sum_{n=0}^{\infty} \Pi_{\gamma}^{(n)} \hbar^{2 n},
\end{equation}
with its coefficient diverging factorially
\begin{equation}
    \Pi_{\gamma}^{(n)} \sim(2 n) !.
\end{equation}
Since the WKB solutions in this subsection are resumed functions, one should regard the Voros symbols and quantum periods as Borel-resummed quantities, but we will omit the notation of Borel resum $\mathcal{S}$.

Until now, we have introduced Stokes graphs which are fully determined by the classical potential $Q_0(x)$, and connection formulas based on the all-order wave function and its Borel summability. With these ingredients, we can analytically continue the WKB solution defined in one Stokes region to another region and rewrite the solutions in one Stokes region as the linear combination of the wave basis for another region. In the content of the WKB analysis, these combination coefficients rely on Voros symbols. In practice, proper boundary conditions are imposed for problems of interest, which give an additional condition for these coefficients. The equality that fulfils boundary conditions is called the exact quantization condition, which is a functional relation in terms of Voros symbols as follows
\begin{equation}
\mathcal{Q}\left(\mathcal{V}_{\gamma_{1}}, \ldots, \mathcal{V}_{\gamma_{r}}\right)=0,
\end{equation}
where $\mathcal{V}_{\gamma_i}$ with $i=1,2,\cdots,r$ are Voros symbols on the relevant one-cycle $\gamma_i$. For example, the EQC for harmonic oscillator can be written as 
\begin{equation}
    1+\mathcal{V}_{\gamma}=0,
\end{equation}
which is the BS quantization condition \eqref{eq:hoeqc}.
We intend to drive the EQCs for the Stark problem in the next section by following the above strategy.

\section{Stark Effect and its quantization conditions}
\label{sc: langer}

Let us write down the Schr\"{o}dinger equation for the Stark effect of hydrogen in a uniform electric field $F$ oriented in the $z$-axis:
\begin{equation}
\label{eq:starkseq}
\left(-\frac{\hbar^{2}}{2}\nabla^{2}-\frac{1}{r}+F z\right)\Psi=E \Psi,
\end{equation}
where atomic units are adopted, and $E$ represents the energy. $r$ denotes the radial coordinate. We keep reduced Planck constant $\hbar$, which is set to $1$ in the numerical calculations of the energy spectrum. 

\subsection{Langer's modification and boundary conditions}

We introduce the  parabolic coordinates $(\xi,\eta,\varphi)$ which are defined by 
\begin{equation}
\label{eq:cor2}
\begin{array}{*{2}{ll}}
&x+i y=\sqrt{\xi \eta}e^{i\varphi}, &\quad \xi=r+z, \\
&z=\frac{1}{2}(\xi-\eta), &\quad \eta=r-z, \\
&r=\frac{1}{2}(\xi+\eta), &\quad \tan \varphi=\frac{y}{x},
\end{array}
\end{equation}
where $r=\sqrt{x^2+y^2+z^2}$, $0\leq\xi,\eta\leq+\infty$ and $0\leqslant\varphi\leqslant2\pi$.
If we consider an ansatz for wave function $\Psi$ as
\begin{equation}
\label{eq:waveansatz}
\Psi=\frac{1}{\sqrt{\xi \eta}}  \psi_{1}(\xi) \psi_{2}(\eta) e^{i m \varphi},
\end{equation}
the three-dimensional Schr\"{o}dinger equation \eqref{eq:starkseq} reduces to a couple of Schr\"{o}dinger-like equations with centrifugal term
\begin{equation}
\begin{aligned}
\label{eq:parabolicseq}
\left(-\hbar^{2} \frac{d^{2}}{d\xi^{2}}+\frac{F}{4}\xi-\frac{E}{2}-\frac{A_{1}}{\xi}+\frac{\hbar^{2} (m^{2}-1)}{4 \xi^{2}}\right) \psi_{1}(\xi)&=0,\\
\left(-\hbar^{2} \frac{d^{2}}{d\eta^{2}}-\frac{F}{4}\eta-\frac{E}{2}-\frac{A_{2}}{\eta}+\frac{\hbar^{2} (m^{2}-1)}{4 \eta^{2}}\right) \psi_{2}(\eta)&=0,
\end{aligned}
\end{equation}
where $A_{1}$ and $A_{2}$ are separation constants satisfying $A_{1}+A_{2}=1$. One can relate the parameters above to those of the potential in the standard form given by \eqref{eq:seq} with 
\begin{equation}
\label{eq:potential}
Q_0(x)=u_0x+u_{1}+\frac{u_{2}}{x}, \quad Q_2(x)=\frac{\ell(\ell+1)}{x^{2}}.
\end{equation}
This enables us to write down the expressions for $Q_0(x)$ in terms of $F$, $E$ and $A_{1,2}$ corresponding to the equations in $\xi$ and $\eta$ coordinates respectively:
\begin{equation}
Q_0(x)=\frac{F}{4} x-\frac{E}{2}-\frac{A_1}{x},\quad \text{or} \quad Q_0(x)=-\frac{F}{4} x-\frac{E}{2}-\frac{A_2}{x}. 
\end{equation}
In both cases, $\ell$ and $m$ are related by 
\begin{equation}
\label{eq:lm}
    \left|\ell+\frac{1}{2}\right|= \left|\frac{m}{2}\right|.
\end{equation}
Equations \eqref{eq:parabolicseq} are suitable for the WKB analysis including all-order expansions but do not provide an accurate leading-order approximation around the origin to the wave function. We use the generalization of the Kramers-Langer substitution \cite{langer3,langer2} by
\begin{equation}
\hbar^2\ell(\ell+1)\to\hbar_i^2(\ell+\frac{1}{2})^2-\frac{\hbar^2}{4},
\end{equation}
or in terms of $m$ as 
\begin{equation}
\label{eq:kramers-langer}
\frac{\hbar^2(m^2-1)}{4}\to\frac{\hbar_i^2m^2}{4}-\frac{\hbar^2}{4}.
\end{equation}
Here another parameter $\hbar_i$ is introduced as an implicit factor, while $\hbar$ is an explicit expansion parameter. The different partitions of the centrifugal term as above represent different expansion schemes but give the same physical solution when $\hbar_i=\hbar$. The WKB expansions for \eqref{eq:kramers-langer} do not match term by term except for the first order and infinite order, see \cite{langer2} for detailed discussion. Since $\hbar_i$ is not involved in power expansions, we can directly set it to $1$ at the beginning. This generalized Langer's modification suggests splitting the centrifugal term into two parts as 
\begin{equation}
\begin{aligned}
\label{eq:parabolicseq2}
\left(-\hbar^{2} \frac{d^{2}}{d\xi^{2}}+\frac{F}{4}\xi-\frac{E}{2}-\frac{A_{1}}{\xi}+\frac{ m^{2}}{4 \xi^{2}}-\frac{\hbar^{2} }{4 \xi^{2}}\right) \psi_{1}(\xi)&=0,\\
\left(-\hbar^{2} \frac{d^{2}}{d\eta^{2}}-\frac{F}{4}\eta-\frac{E}{2}-\frac{A_{2}}{\eta}+\frac{ m^{2}}{4 \eta^{2}}-\frac{\hbar^{2}}{4 \eta^{2}}\right) \psi_{2}(\eta)&=0.
\end{aligned}
\end{equation}
Now we can set 
\begin{equation}
\label{eq:qlangerxi}
    Q_0(\xi)=\frac{F}{4}\xi-\frac{E}{2}-\frac{A_{1}}{\xi}+\frac{ m^{2}}{4 \xi^{2}}, \quad Q_2(\xi)=-\frac{1}{4 \xi^{2}},
\end{equation}
for $\xi$ coordinate and 
\begin{equation}
\label{eq:qlangereta}
    Q_0(\eta)=-\frac{F}{4}\eta-\frac{E}{2}-\frac{A_{2}}{\eta}+\frac{ m^{2}}{4 \eta^{2}}, \quad Q_2(\eta)=-\frac{1}{4 \eta^{2}},
\end{equation}
for $\eta$ coordinate. Let us then check the behaviours of the wave functions at the boundaries. It is enough to take account of the leading order WKB solution of \eqref{eq:wkb} at this stage
 \begin{equation}
 \label{eq:wkbapp}
    	\psi(x)\sim \frac{1}{\sqrt{p_0(x)}}\exp\left( \pm\frac{1}{\hbar}\int_{a}^{x}p_{0}(x)\rd x\right).
\end{equation}
We temporarily set $\hbar=1$ for simplicity, and consider the $\xi$-equation at first. Near the origin, the inverse square term dominates the wave function
\begin{equation}
    Q_0(\xi)\sim\frac{m^2}{4\xi^2},\quad p_0(\xi)\sim\frac{|m|}{2}\frac{1}{\xi}.
\end{equation}
Then the wave function near the origin behaves like
\begin{equation}
\label{eq:boundary1}
\psi_{1}^{\pm}(\xi)\sim
    \frac{1}{\sqrt{p_0(\xi)}}\exp(\pm\int^\xi p_0(\xi)d\xi)\sim\xi^{\frac{1}{2}\pm\frac{|m|}{2}}\sim\begin{cases}
        \xi^{\ell+1}, &\text{for}\,           +,\\
        \xi^{-\ell}, &\text{for}\,  -,
    \end{cases}
\end{equation}
where \eqref{eq:lm} is used at the last step. We then obtain one normalizable solution $\psi_{1}^{+}(\xi)$, that is proper for the Stark problem, and one divergent solution $\psi_{1}^{-}(\xi)$. Now let us turn to consider the behaviour of the solution at infinity, the dominant contribution to $p_0(\xi)$ is
\begin{equation}
    Q_0(\xi)\sim\frac{F}{4}\xi,\quad p_0(\xi)\sim\frac{F}{2}\xi^{\frac{1}{2}}.
\end{equation}
It is easy to find the following asymptotic approximation 
\begin{equation}
\label{eq:boundary2}
    \psi_{1}^{\pm}(\xi)\sim\xi^{-\frac{1}{4}}e^{\pm\frac{F}{3}\xi^{\frac{3}{2}}}.
\end{equation}
We get one increasing solution and one exponentially suppressed solution, the latter is what we expect, namely $\psi_{1}^{-}(\xi)$ at the asymptotic infinite region. 
The wave function for the $\eta$-equation can be analyzed in the same way. Near the origin, it is entirely the same as that for the $\xi$-equation. 
\begin{equation}
\label{eq:boundary3}
\psi_{2}^{\pm}(\eta)\sim
    \frac{1}{\sqrt{p_0(\eta)}}\exp(\pm\int^\eta p_0(\eta)d\eta)\sim\eta^{\frac{1}{2}\pm\frac{|m|}{2}}\sim\begin{cases}
        \eta^{\ell+1}, &\text{for}\,           +,\\
        \eta^{-\ell}, &\text{for}\,-.
    \end{cases}
\end{equation}
The boundary condition requires the $\psi_{2}^{+}$ component to survive.
At large $\eta$ region, the dominant part for $p_0(\eta)$ changes to 
\begin{equation}
    Q_0(\eta)\sim-\frac{F}{4}\eta,\quad p_0(\eta)\sim i\frac{F}{2}\eta^{\frac{1}{2}}, 
\end{equation}
then the wave function asymptotically behaves like
\begin{equation}
\label{eq:boundary4}
    \psi_{2}^{\pm}(\eta)\sim\eta^{-\frac{1}{4}}e^{\pm i\frac{F}{3}\eta^{\frac{3}{2}}}.
\end{equation}
This time we obtain one outgoing wave $\psi_{2}^{+}(\eta)$ and one ingoing wave $\psi_{2}^{-}(\eta)$, the former satisfies the required boundary condition. 

Then we are going to apply connection formulas in section \ref{sec:wkb} to relate the wave functions from one Stokes region to another region and impose the above boundary conditions, which generates a functional relation for the quantum periods and gives us exact quantization conditions for the Stark problem.
\subsection{Quantization conditions}

The quantization conditions can be derived from the Stokes graphs incorporating boundary conditions discussed in the last section. For equations \eqref{eq:parabolicseq2}, there are three turning points $a$, $b$, and $c$ when $m\neq0$. As for the $m=0$ case, we will deal with it separately in the later calculation. One can plot the Stokes lines starting from each turning point by solving condition \eqref{eq:stokescurve}. The Stokes graph can be established by combining all the Stokes lines in the complex $\xi$ or $\eta$ coordinate plane.

\begin{figure}[H]
\centering
    \subfigure[Distinct regions for the $\xi$-equation]{
        \begin{minipage}{0.47\linewidth}
         \centering
    \includegraphics[width=1.0\linewidth]{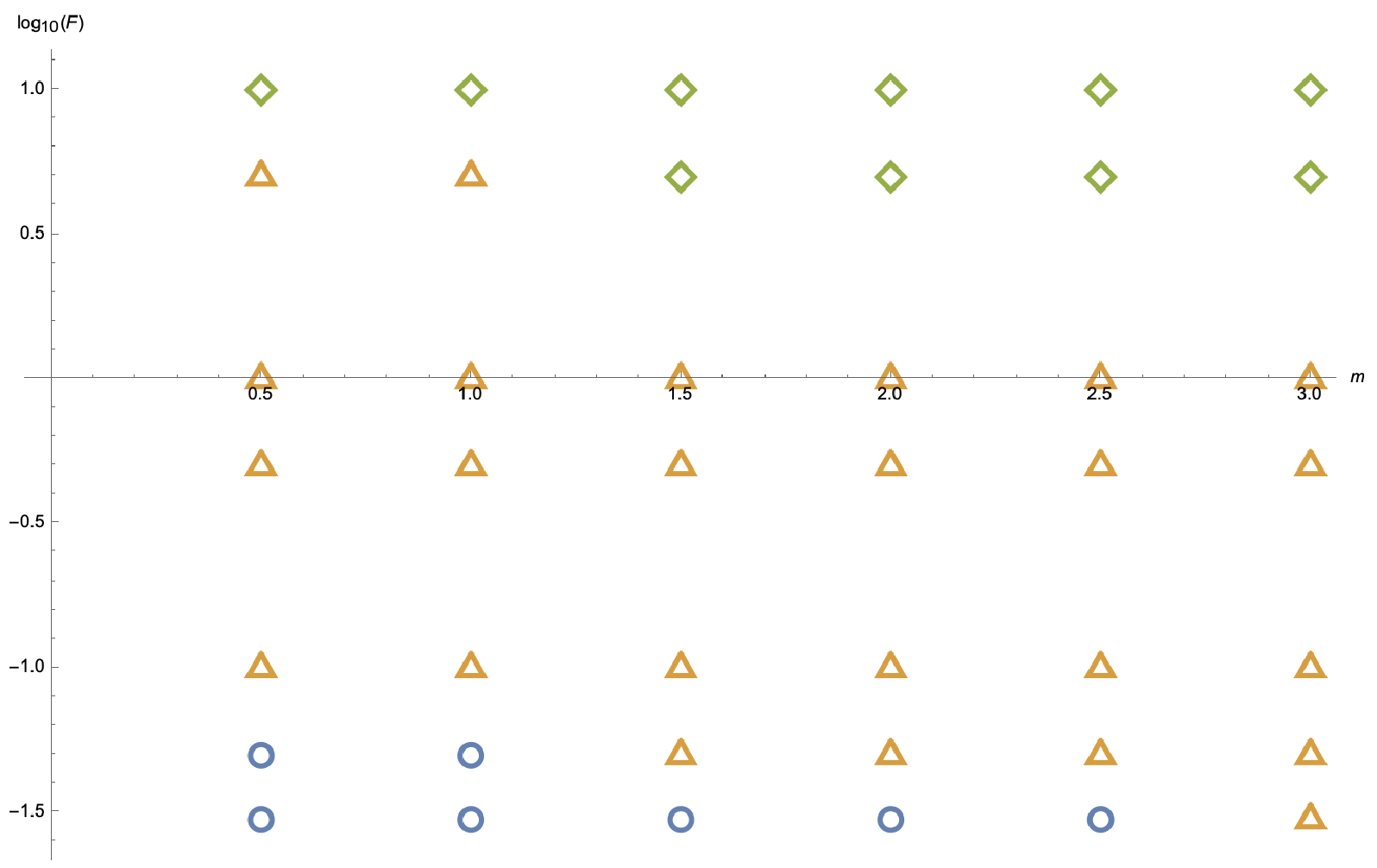}
    \label{fig:transition1}
    \end{minipage}
    }
     \subfigure[Distinct regions for the $\eta$-equation]{
        \begin{minipage}{0.47\linewidth}
         \centering
    \includegraphics[width=1.0\linewidth]{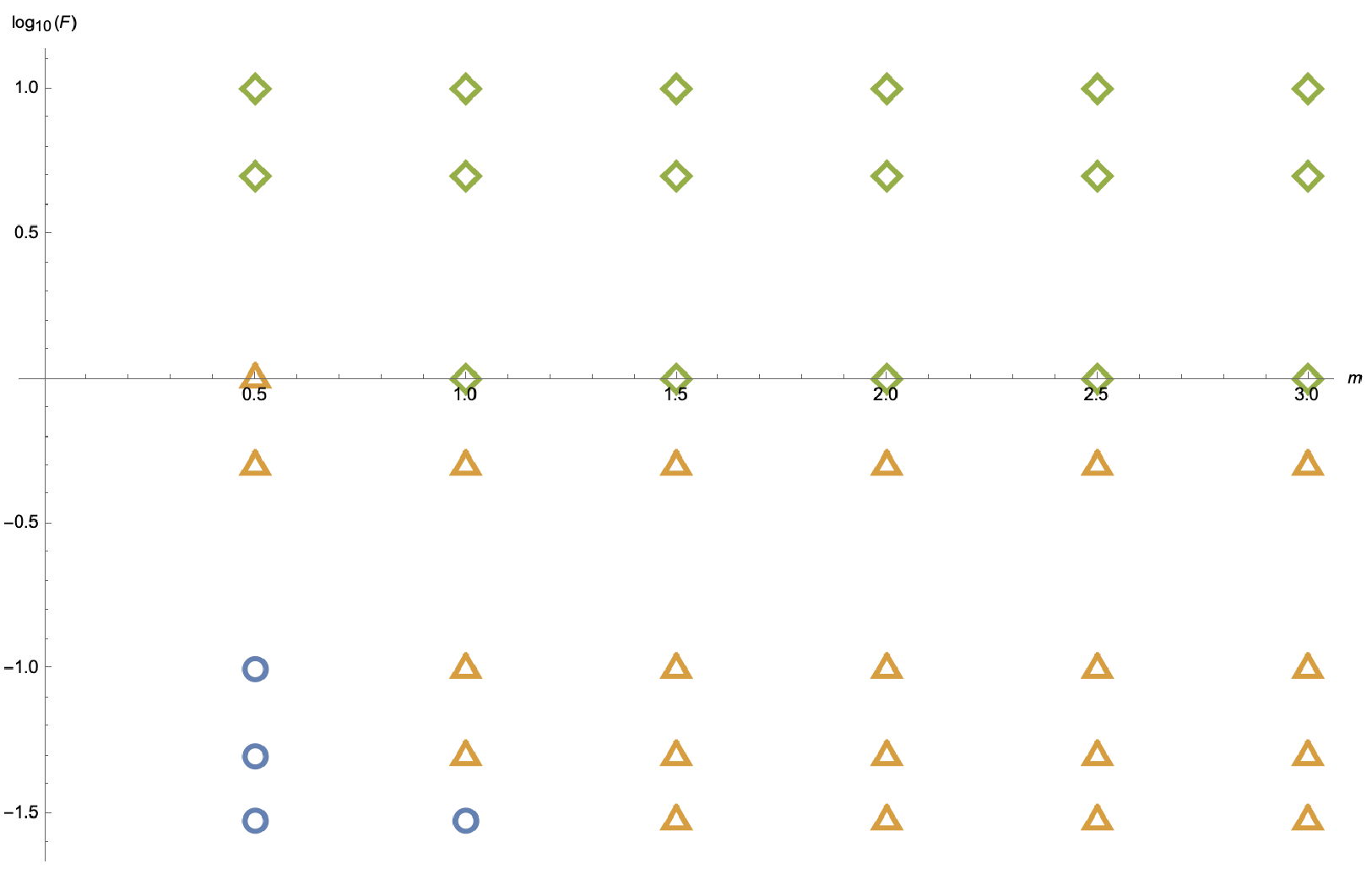}
     \label{fig:transition2}
    \end{minipage}
    }
    \caption{The left and right figures show the three distinct regions of parameters $(F, m)$ for $\xi$ and $\eta$-equations respectively according to the Stokes graphs. We will denote these three regions, 
$\mathrm{I}, \mathrm{II},\mathrm{III}$ from small $F$ to large $F$ respectively for both equations.
$m$ takes values from $\frac{1}{2}$ to $3$ with step of half-integer, while $F=0.03, 0.05, 0.1, 0.5, 1, 5, 10$. We use the logarithmic axis for $F$.}
    \label{fig: transition}
\end{figure}

We find the interesting phenomenon that the Stokes graph changes its topology when changing the values of $F$ and $m$. These are three distinct regions of parameters in the $(F, m)$ plane for both equations, see figure \ref{fig: transition}, and each of them corresponds to a special configuration of the Stokes graphs as graphically shown in figure \ref{fig:starkstokes}. These regions are delimited by the curves in the $(F, m)$ plane characterized by
\begin{equation}
\label{eq: transition}
    \operatorname{Im}\frac{1}{\hbar}\int_{b}^{c} \sqrt{Q_0(\xi)}d\xi=0,\quad  \operatorname{Im}\frac{1}{\hbar}\int_{a}^{c} \sqrt{Q_0(\xi)}d\xi=0,
\end{equation}
for $\xi$-equation and 
\begin{equation}
\label{eq: transition2}
    \operatorname{Im}\frac{1}{\hbar}\int_{b}^{c} \sqrt{Q_0(\eta)}d\eta=0,
\end{equation}
for $\eta$-equation. Once the above conditions are satisfied, one Stokes line connects two turning points, and the Stokes graph becomes degenerate. These conditions are involved, since $E, A_{1,2}$ depend on $F$ and $m$, and are determined with $F$ and $m$ specified. There is no simple way to determine the delimited curves explicitly to our knowledge. In principle, one can figure out the corresponding Stokes graph for specified $F$ and $m$ with $E$ and $A_{1,2}$ determined by other methods to confirm which region it belongs to. We use $F_{\xi,1}(m)$ and $F_{\xi,2}(m)$ to denote the first and second critical values of $F$ for a specified $m$, and the similar notations $F_{\eta,1}(m)$ and $F_{\eta,2}(m)$ for $\eta$-equation.

Let us now focus on the Stokes graph for the $\xi$-equation in the region $\mathrm{I}_\xi$, its Stokes graph is shown in figure 3.\ref{fig:fig1}. To derive the quantization condition, we have to know the relation of wave functions between origin and infinity, which can be accomplished by using connection formulas in section \ref{sec:wkb}. One can continue the solutions at infinity to the origin along the green arrow depicted in figure 3.\ref{fig:fig1}, this procedure can be written as follows.

\begin{enumerate}
\item We start from the wave function basis in Stokes region $\mathrm{I}_{\infty}$ normalized at $a$, it is necessary at first to cross the branch cut which exchanges the exponentially suppressed and growing solutions, this progress can be written as
\begin{equation}
\left(\begin{array}{l}
\iota \psi_{a, \mathrm{I}_\infty}^{-}(\xi) \\
\iota \psi_{a, \mathrm{I}_\infty}^{+}(\xi)
\end{array}\right)= \left(\begin{array}{l}
\psi_{a, \mathrm{I}_\infty}^{+}(\xi) \\
\psi_{a, \mathrm{I}_\infty}^{-}(\xi)
\end{array}\right),
\end{equation}
where the operator $\iota$ maps the function in one square sheet to one in another sheet.
\item Now we are working in the same sheet, we continue the wave functions $\psi_{a, \mathrm{I}_\infty}^{\pm}(\xi)$ to cross the Stokes line labelled by $+$ anticlockwise to get basis in the intermediate region $\mathrm{I}_m$ normalized at the same turning point $a$ by the multiplication of the monodromy matrix $M_+$
\begin{equation}
\left(\begin{array}{l}
\psi_{a, \mathrm{I}_\infty}^{+}(\xi) \\
\psi_{a, \mathrm{I}_\infty}^{-}(\xi)
\end{array}\right)=M_{+}\left(\begin{array}{l}
\psi_{a, \mathrm{I}_m}^{+}(\xi) \\
\psi_{a, \mathrm{I}_m}^{-}(\xi)
\end{array}\right).
\end{equation}
\item We obtain the basis in the region $\mathrm{I}_m$ normalized at $a$, to continue the procedure along the green line. It is necessary to arrive at the basis normalized at $b$ for the same Stokes region $\mathrm{I}_m$. It is accomplished by the normalization matrix
\begin{equation}
\left(\begin{array}{l}
\psi_{a, \mathrm{I}_m}^{+}(\xi) \\
\psi_{a, \mathrm{I}_m}^{-}(\xi)
\end{array}\right)=N_{a b}\left(\begin{array}{l}
\psi_{b, \mathrm{I}_m}^{+}(\xi) \\
\psi_{b, \mathrm{I}_m}^{-}(\xi)
\end{array}\right).
\end{equation}
\item As the same as the second step, by crossing a Stokes line labelled by $+$ anticlockwise, one gets the basis in the region $\mathrm{I}_0$ normalized at $b$
\begin{equation}
\left(\begin{array}{l}
\psi_{b, \mathrm{I}_m}^{+}(\xi) \\
\psi_{b, \mathrm{I}_m}^{-}(\xi)
\end{array}\right)=M_+\left(\begin{array}{l}
\psi_{b, \mathrm{I}_0}^{+}(\xi) \\
\psi_{b, \mathrm{I}_0}^{-}(\xi)
\end{array}\right).
\end{equation}
\item We finally pull back the normalization from $b$ to $a$ for comparison.
\begin{equation}
\left(\begin{array}{l}
\psi_{b, \mathrm{I}_0}^{+}(\xi) \\
\psi_{b, \mathrm{I}_0}^{-}(\xi)
\end{array}\right)=N_{b a}\left(\begin{array}{l}
\psi_{a, \mathrm{I}_0}^{+}(\xi) \\
\psi_{a, \mathrm{I}_0}^{-}(\xi)
\end{array}\right).
\end{equation}
\end{enumerate}
Accounting for all these transformations, the two bases are related as
\begin{equation}
\left(\begin{array}{l}
\iota \psi_{a, \mathrm{I}_\infty}^{-}(\xi) \\
\iota \psi_{a, \mathrm{I}_\infty}^{+}(\xi)
\end{array}\right)=M_{+} N_{a b} M_{+} N_{b a}\left(\begin{array}{l}
\psi_{a, \mathrm{I}_0}^{+}(\xi) \\
\psi_{a, \mathrm{I}_0}^{-}(\xi)
\end{array}\right)
=\left(\begin{array}{c}
\psi_{a, \mathrm{I}_0}^{+}(\xi)+i(1+\mathcal{V}_{\gamma}) \psi_{a, \mathrm{I}_0}^{-}(\xi) \\
\psi_{a, \mathrm{I}_0}^{-}(\xi)
\end{array}\right),
\end{equation}
where $\mathcal{V}_{\gamma}$ is the Voros symbol, and $\gamma$ is the trajectory enclosing turning points $b$ and $a$. This expression relates the wave function basis at positive infinity  to that around the origin. We note that this formula is very similar to the Q-function in the ODE/IM correspondence \cite{DT} and also the numerical method used in \cite{Damburg1976,Kolosov83}. 
The boundary conditions for \eqref{eq:boundary1} and \eqref{eq:boundary2} require  $\psi_{a,\mathrm{I}_0}^{-}$ and $\iota \psi_{a,\mathrm{I}_\infty}^{+}$ to vanish, which deduce the following quantization condition
\be
\mathcal{Q}(\mathcal{V}_{\gamma}):=1+\mathcal{V}_{\gamma}=1+\exp\left(\frac{1}{\hbar}\oint_{\gamma}P_{\mathrm{even}}(\xi)d \xi\right)=0.
\ee
We can then write down the Bohr-Sommerfeld-type quantization condition
\be
\label{eq:qc1}
\Pi_\gamma(\hbar)=\oint_\gamma P_{\mathrm{even}}(\xi) d \xi=2 \pi i \hbar\left(n_\xi+\frac{1}{2}\right),\, n_\xi=0,1,2,\cdots,\, \text{for region}\, \mathrm{I}_\xi,
\ee
where $n_{\xi}$ denotes the quantum number corresponding to $\xi$ coordinate. Let us emphasize that the quantum periods should be understood as their Borel resummation. This quantization condition was presented for the Stark problem a long time ago such as \cite{starkwkb82}, where the leading order contribution is considered. Here we derive it rigorously and show it is exactly satisfied when considering quantum corrections.

\begin{figure}[H]
\centering
    \subfigure[Stokes graph for $\xi$-equation in region $\mathrm{I}_\xi$]{
        \begin{minipage}{0.45\linewidth}
         \centering
    \includegraphics[width=0.90\linewidth]{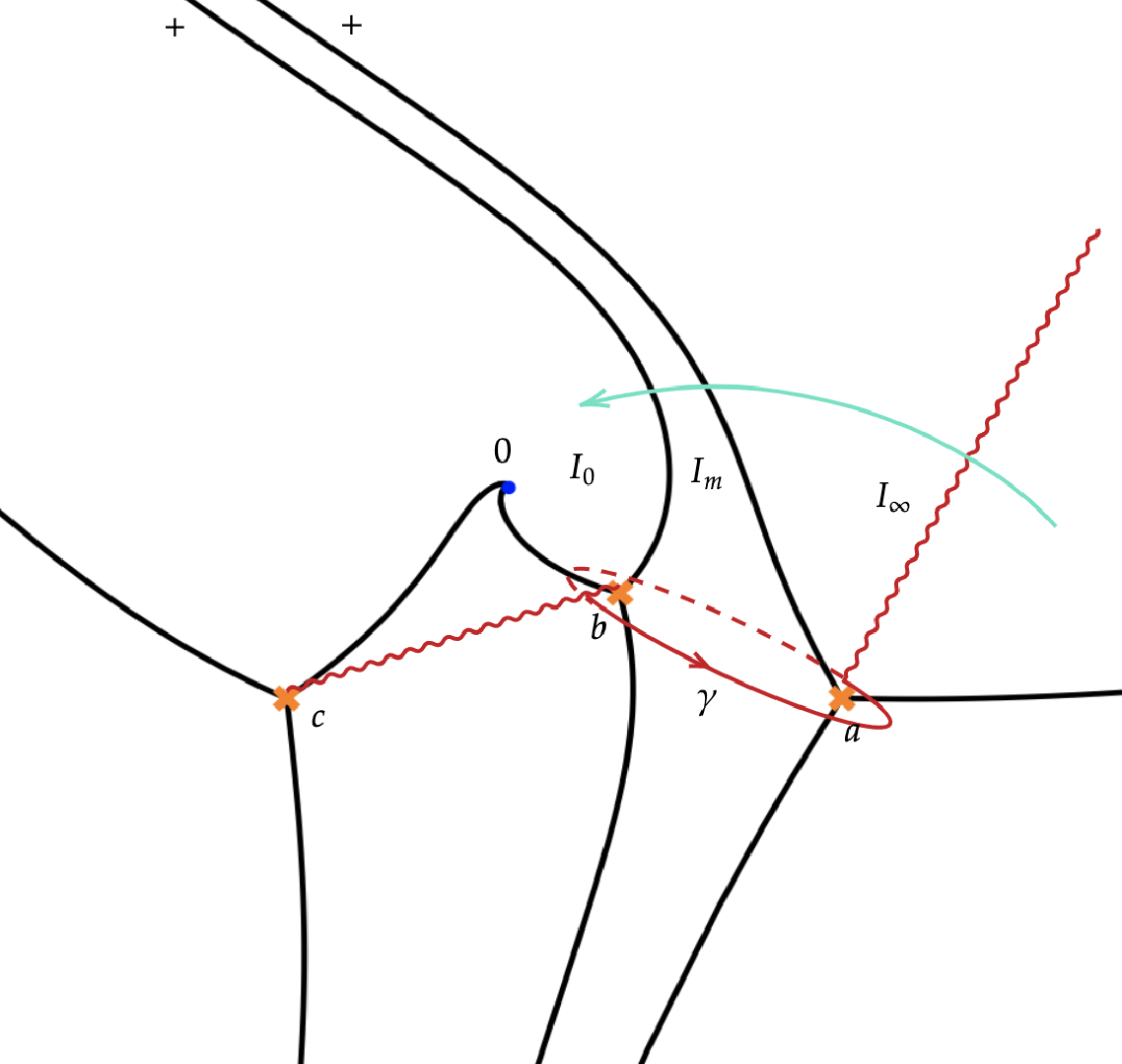}
    \label{fig:fig1}
    \end{minipage}
    }
     \subfigure[Stokes graph for $\eta$-equation in region $\mathrm{I}_\eta$]{
        \begin{minipage}{0.45\linewidth}
         \centering
    \includegraphics[width=0.90\linewidth]{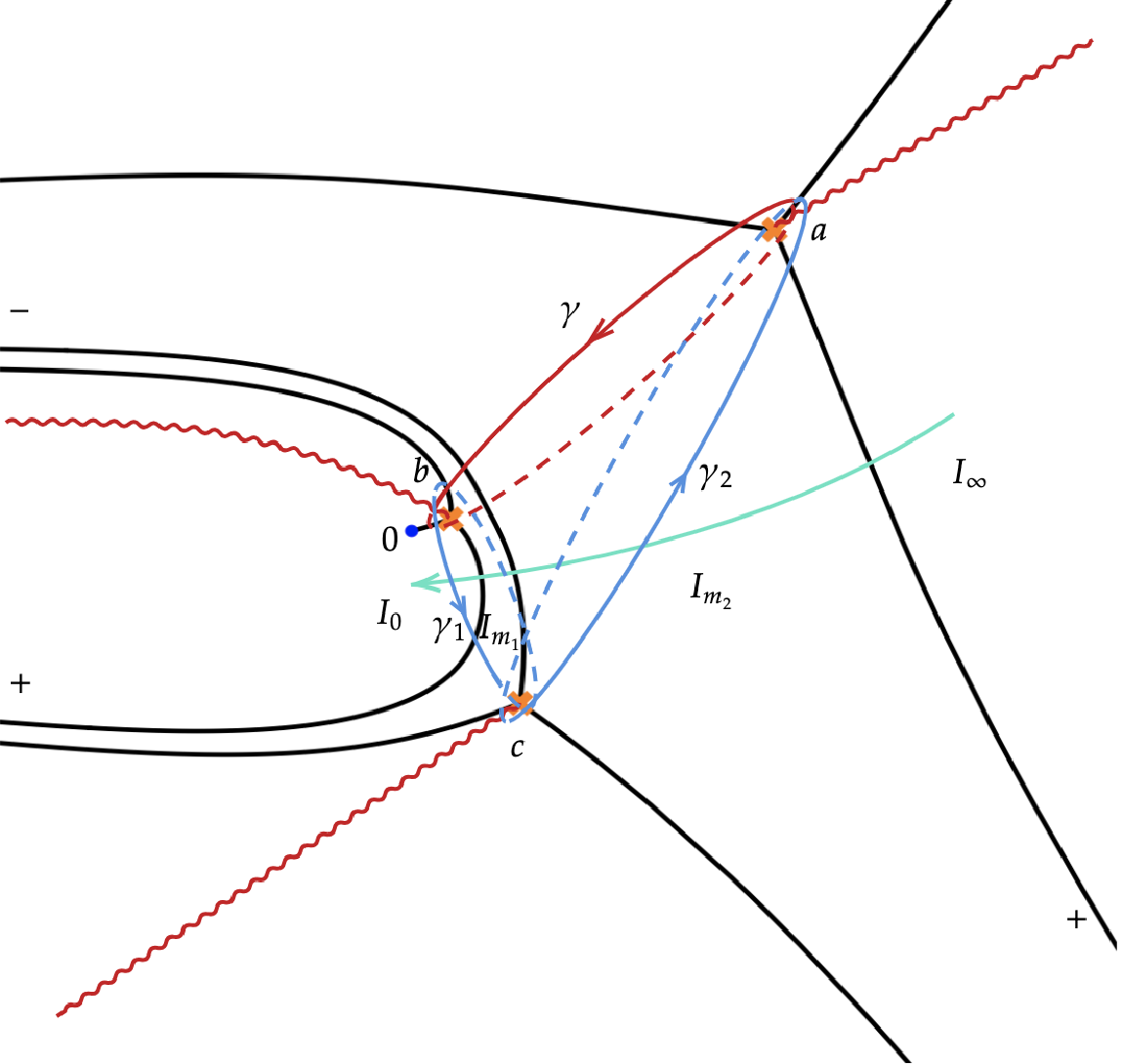}
     \label{fig:fig0}
    \end{minipage}
    }

      \subfigure[Stokes graph for $\xi$-equation in  region $\mathrm{II}_\xi$]{
        \begin{minipage}{0.45\linewidth}
         \centering
    \includegraphics[width=0.90\linewidth]{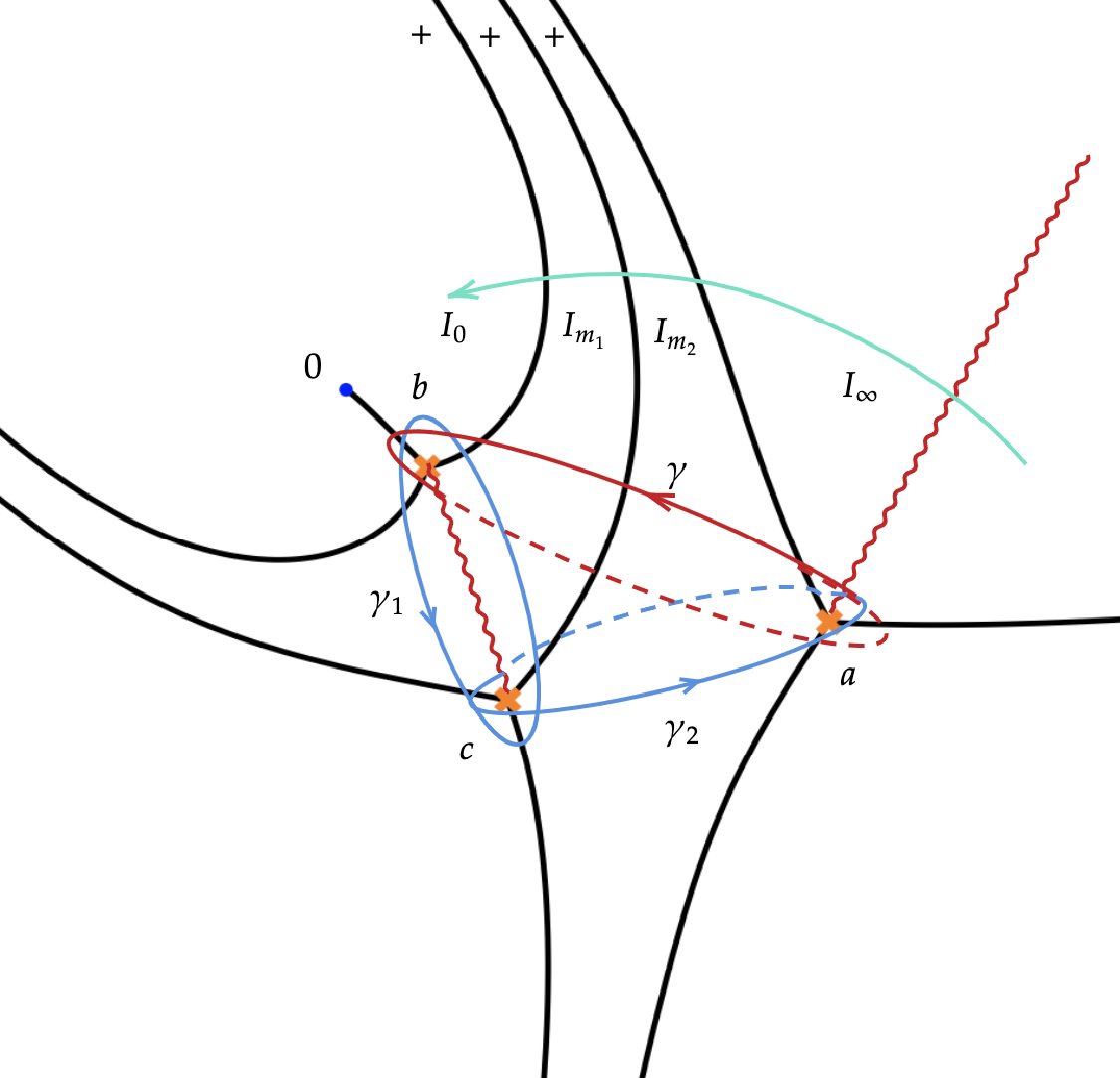}
     \label{fig:fig2}
    \end{minipage}
    }
     \subfigure[Stokes graph for $\eta$-equation in region $\mathrm{II}_\eta$]{
        \begin{minipage}{0.45\linewidth}
         \centering
    \includegraphics[width=0.90\linewidth]{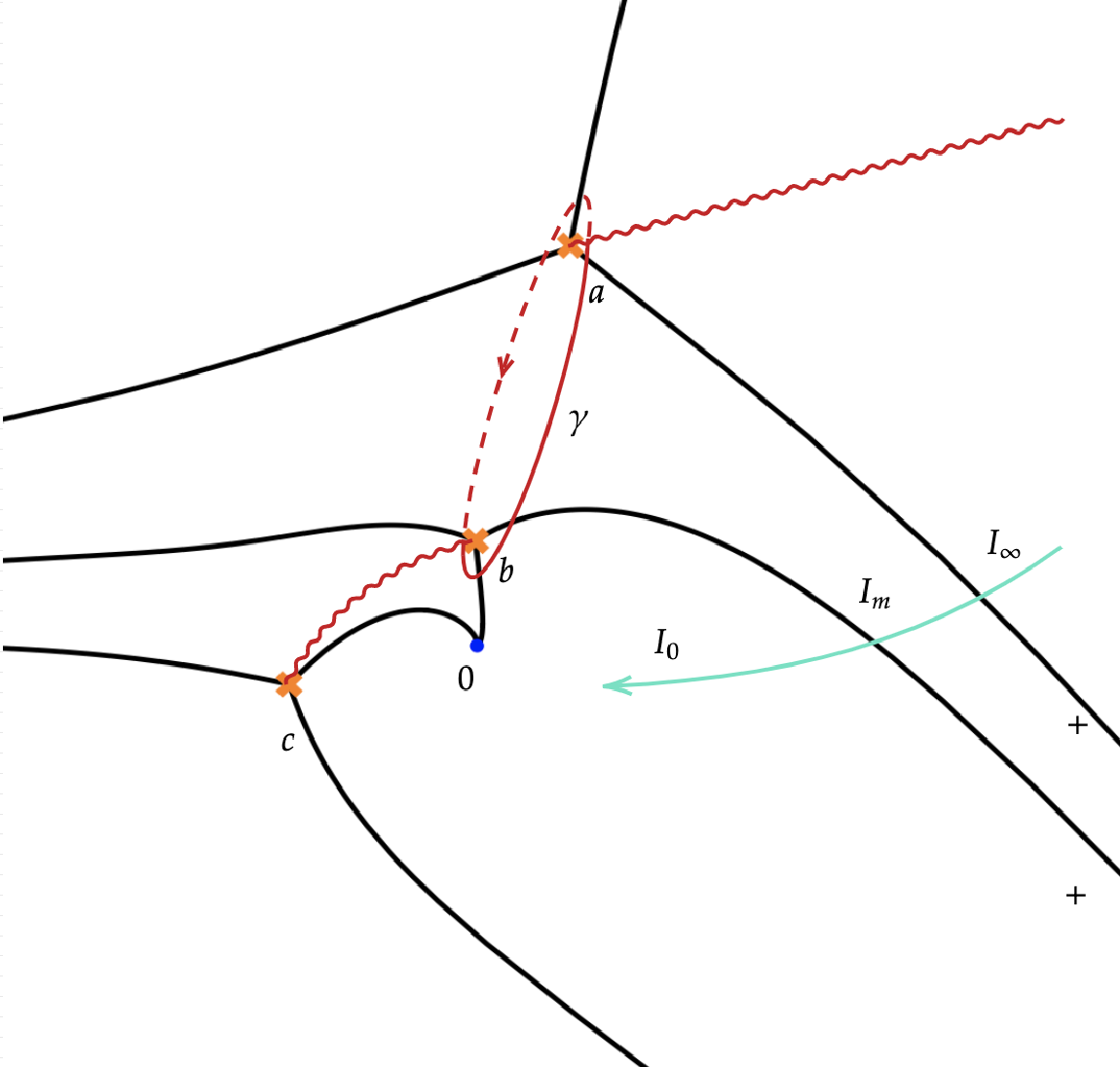}
    \label{fig:fig3}
    \end{minipage}
    }
\ContinuedFloat
\end{figure}
\begin{figure}[H]
       \subfigure[Stokes graph for $\xi$-equation in  region $\mathrm{III}_\xi$]{
        \begin{minipage}{0.45\linewidth}
         \centering
    \includegraphics[width=0.90\linewidth]{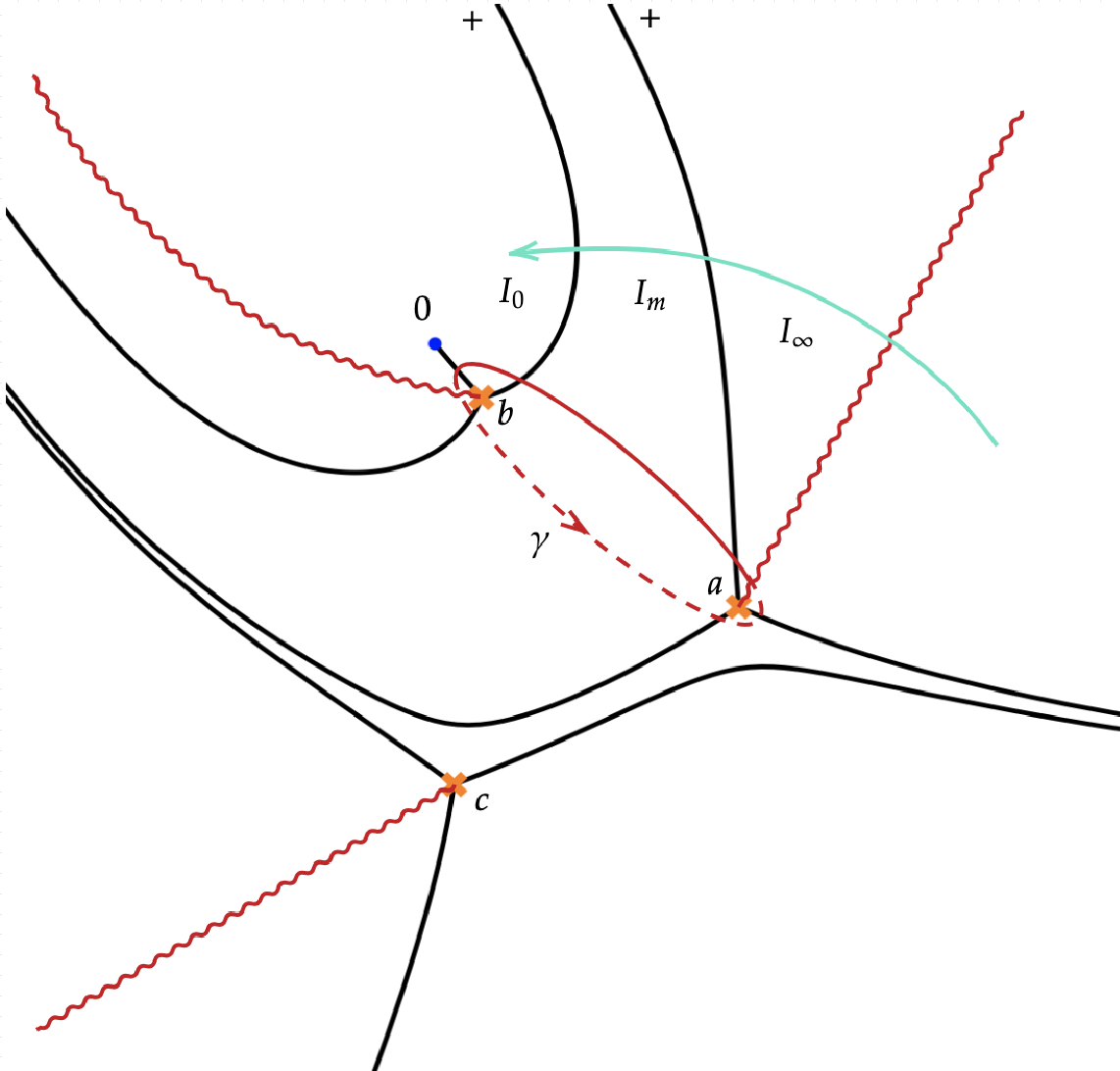}
     \label{fig:fig5}
    \end{minipage}
    }
     \subfigure[Stokes graph for $\eta$-equation in region $\mathrm{III}_\eta$]{
        \begin{minipage}{0.45\linewidth}
         \centering
    \includegraphics[width=0.90\linewidth]{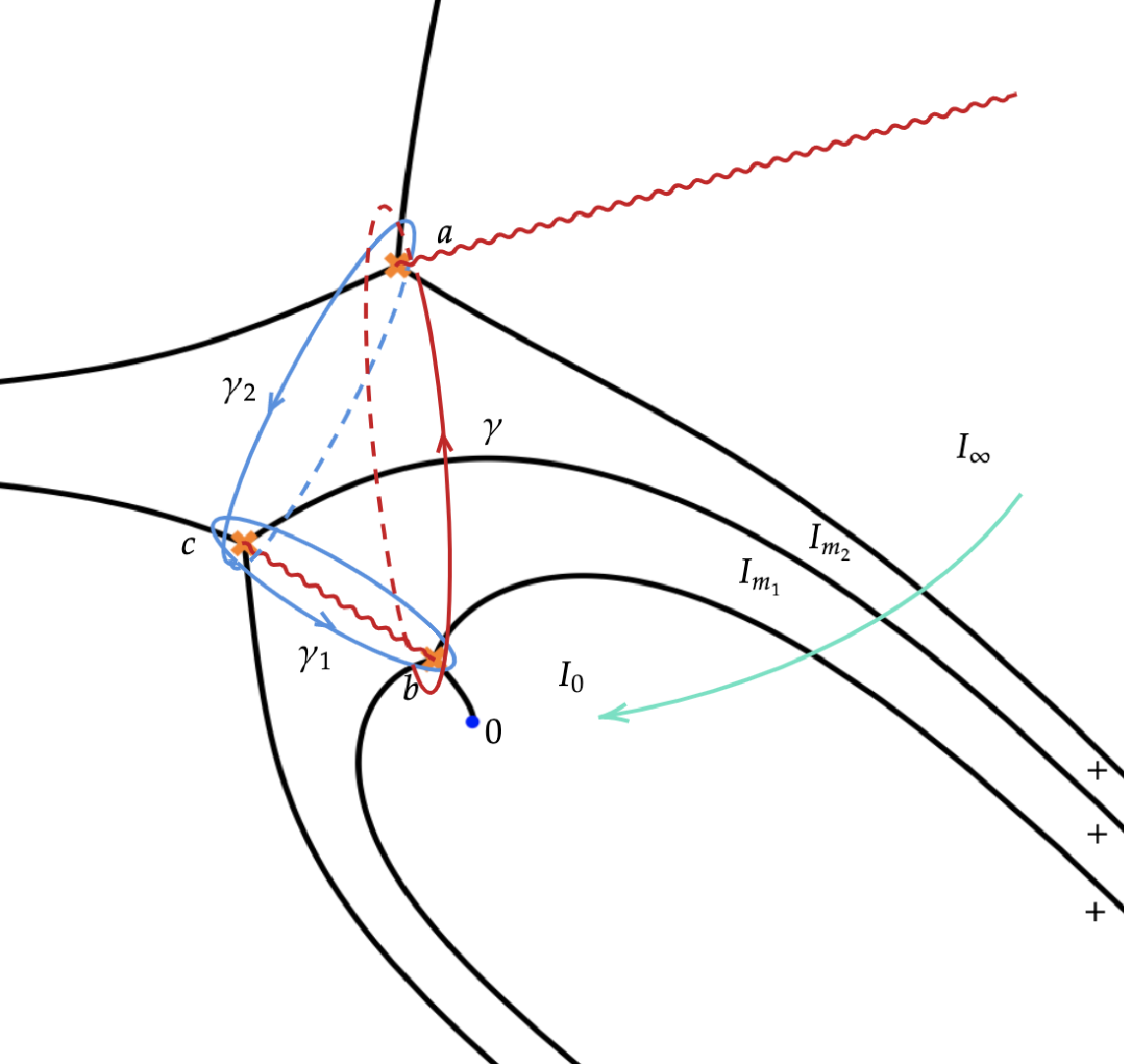}
     \label{fig:fig4}
    \end{minipage}
    }
    \caption{Stokes graphs and its transition. The orange cross denotes three turning points $a$, $b$, and $c$ respectively. The blue point is the origin. The red wavy lines are branch cuts and the green arrow represents the continuation path. $\gamma, \gamma_1$, and $\gamma_2$ are one-cycle encircling turning points, dashed lines indicate the contour enters the second sheet after crossing a branch cut. The Stokes lines are represented by black lines with $+$ and $-$ indicating the orientation. $\mathrm{I}_{0}, \mathrm{I}_{m}$, and $\mathrm{I}_{\infty}$ are adjacent Stokes regions.}
    \label{fig:starkstokes}
\end{figure}

Then, we turn our attention to the Stokes graph for $\xi$-equation in regions $\mathrm{II}_\xi$, as shown in figure 3.\ref{fig:fig2}. We follow a similar process as discussed before. The connection of the solutions from infinity to the origin for 3.\ref{fig:fig2} is given by the matrix multiplication
\begin{equation}
\begin{aligned}
\left(\begin{array}{l}
\iota \psi_{a, \mathrm{I}_\infty}^{-}(\xi) \\
\iota \psi_{a, \mathrm{I}_\infty}^{+}(\xi)
\end{array}\right)&=M_{+} N_{a c} M_{+} N_{c b}M_{+}N_{b c}N_{c a}\left(\begin{array}{l}
\psi_{a, \mathrm{I}_0}^{+}(\xi) \\
\psi_{a, \mathrm{I}_0}^{-}(\xi)
\end{array}\right)\\
&=\left(\begin{array}{c}
\psi_{a, \mathrm{I}_0}^{+}(\xi)+i(1+\mathcal{V}_{\gamma_2}+\mathcal{V}_{\gamma_1}\mathcal{V}_{\gamma_2}) \psi_{a, \mathrm{I}_0}^{-}(\xi) \\
\psi_{a, \mathrm{I}_0}^{-}(\xi)
\end{array}\right),
\end{aligned}
\end{equation}
Then the boundary condition requires
\be
\label{eq:eqc2}
\mathcal{Q}(\mathcal{V}_{\gamma},\mathcal{V}_{\gamma_1}):=1+\mathcal{V}_{\gamma}+\mathcal{V}_{\gamma_2}=1+\mathcal{V}_{\gamma}\left(1+\mathcal{V}_{\gamma_1}^{-1}\right)=0,
\ee
which is the exact quantization condition for the region $\mathrm{II}_\xi$.
In the last equality, we use the fact 
\begin{equation}
\Pi_{\gamma_1}+\Pi_{\gamma_2}=\Pi_{\gamma} \quad \text{or} \quad \mathcal{V}_{\gamma_1}\mathcal{V}_{\gamma_2}=\mathcal{V}_{\gamma}.
\end{equation}
We can write the quantization condition of the modified Bohr-Sommerfeld type from \eqref{eq:eqc2} as 
\begin{equation}
\label{eq:qc2}
\Pi_\gamma(\hbar)+\log\left(1+e^{-\frac{1}{\hbar}\Pi_{\gamma_{1}}(\hbar)}\right)=2 \pi i \hbar\left(n_\xi+\frac{1}{2}\right),\, n_\xi=0,1,2,\cdots,\, \text{for region}\, \mathrm{II}_\xi.
\end{equation}
Here $\Pi_{\gamma}(\hbar)$ and $\Pi_{\gamma_1}(\hbar)$ are also Borel-resummed quantities. It was found that $\mathcal{V}_{\gamma_1}$ gives an exponentially suppressed contribution, so the BS quantization condition still gives an approximation in the small $\hbar$ limit. The additional logarithmic term plays a role in non-perturbative correction. This becomes readily apparent by expanding the logarithmic term.

It is obvious that the connection problem for region $\mathrm{III}_\xi$
are the same with that for region $\mathrm{I}_\xi$ from the configuration of the Stokes graphs, figure 3.\ref{fig:fig1} and 3.\ref{fig:fig5}. So we can directly write down the exact quantization condition 
\be
\label{eq:qc5}
\Pi_\gamma(\hbar)=\oint_\gamma P_{\mathrm{even}}(\xi) d \xi=2 \pi i \hbar\left(n_\xi+\frac{1}{2}\right),\, n_\xi=0,1,2,\cdots,\, \text{for region}\, \mathrm{III}_\xi.
\ee

\begin{figure}[ht]
\centering
    \subfigure[$F<F_{\xi,1}(1)$, $\theta=0$]{
        \begin{minipage}{0.4\linewidth}
         \centering
    \includegraphics[width=0.8\linewidth]{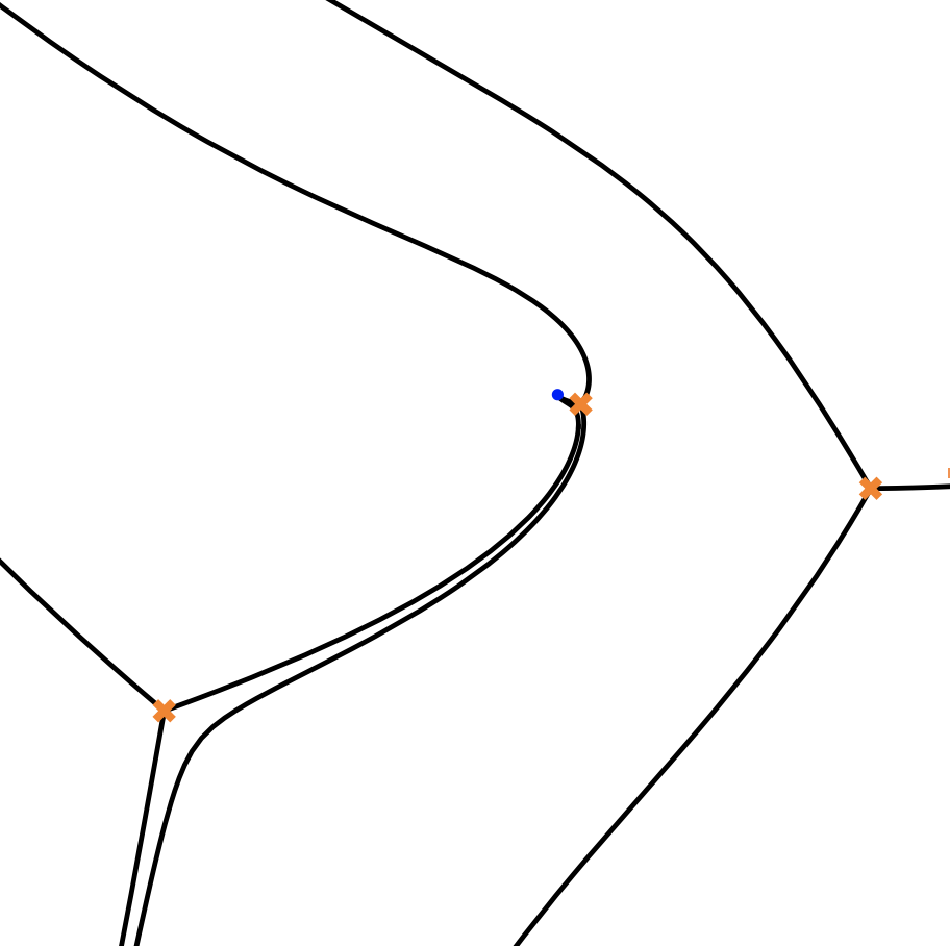}
    \label{fig:fig21}
    \end{minipage}
    }
     \subfigure[$F<F_{\xi,1}(1)$, $\theta=\delta>0$]{
        \begin{minipage}{0.4\linewidth}
         \centering
    \includegraphics[width=0.8\linewidth]{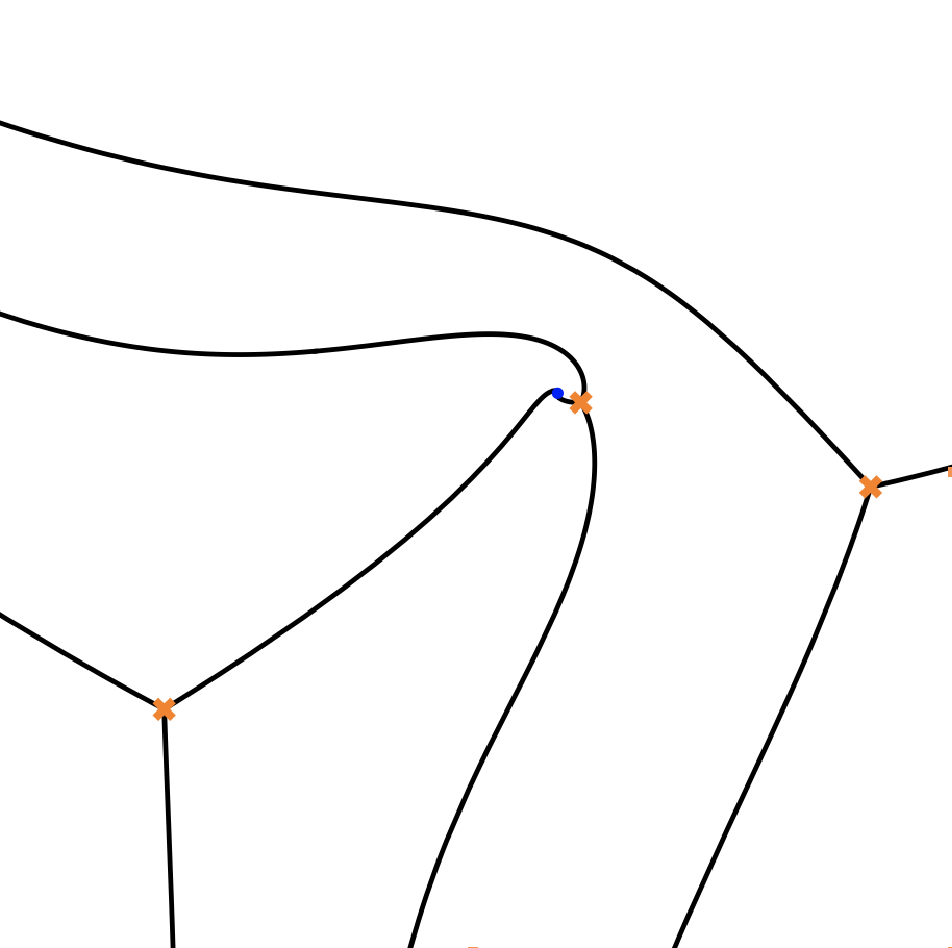}
     \label{fig:fig22}
    \end{minipage}
    }

     \subfigure[$F>F_{\xi,1}(1)$, $\theta=\delta>0$]{
        \begin{minipage}{0.4\linewidth}
         \centering
    \includegraphics[width=0.8\linewidth]{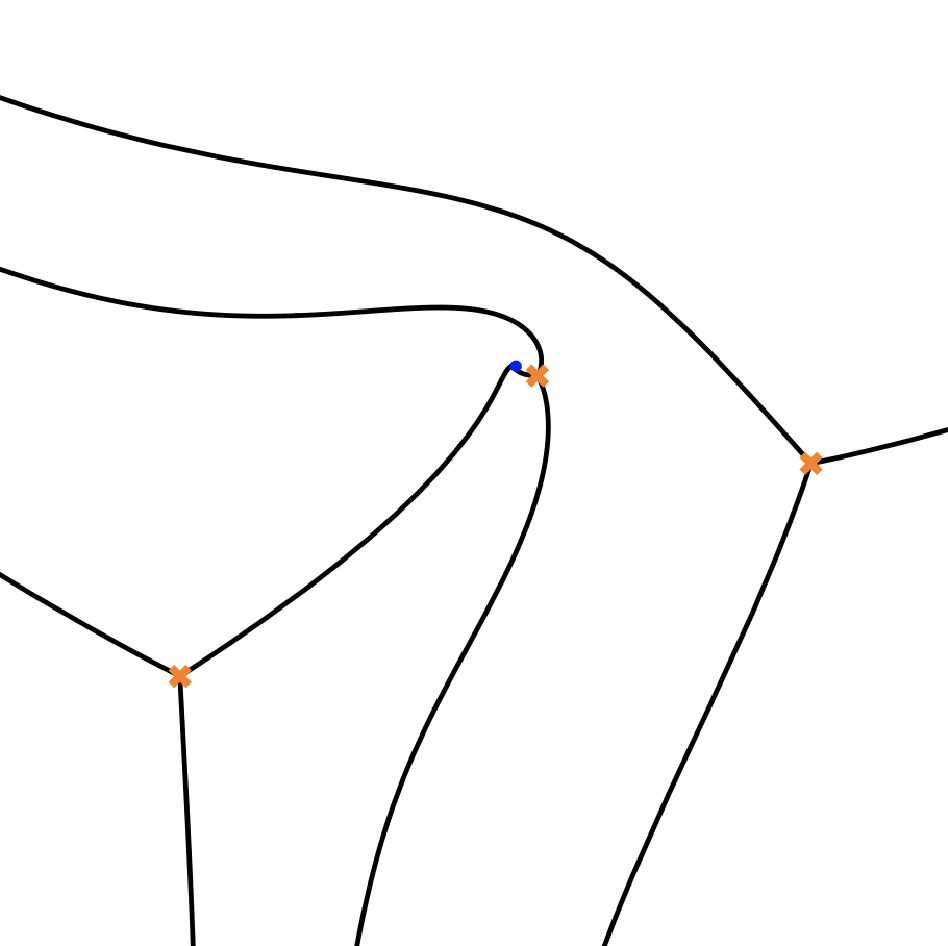}
    \label{fig:fig23}
    \end{minipage}
    }
     \subfigure[$F>F_{\xi,1}(1)$, $\theta=0$]{
        \begin{minipage}{0.4\linewidth}
         \centering
    \includegraphics[width=0.8\linewidth]{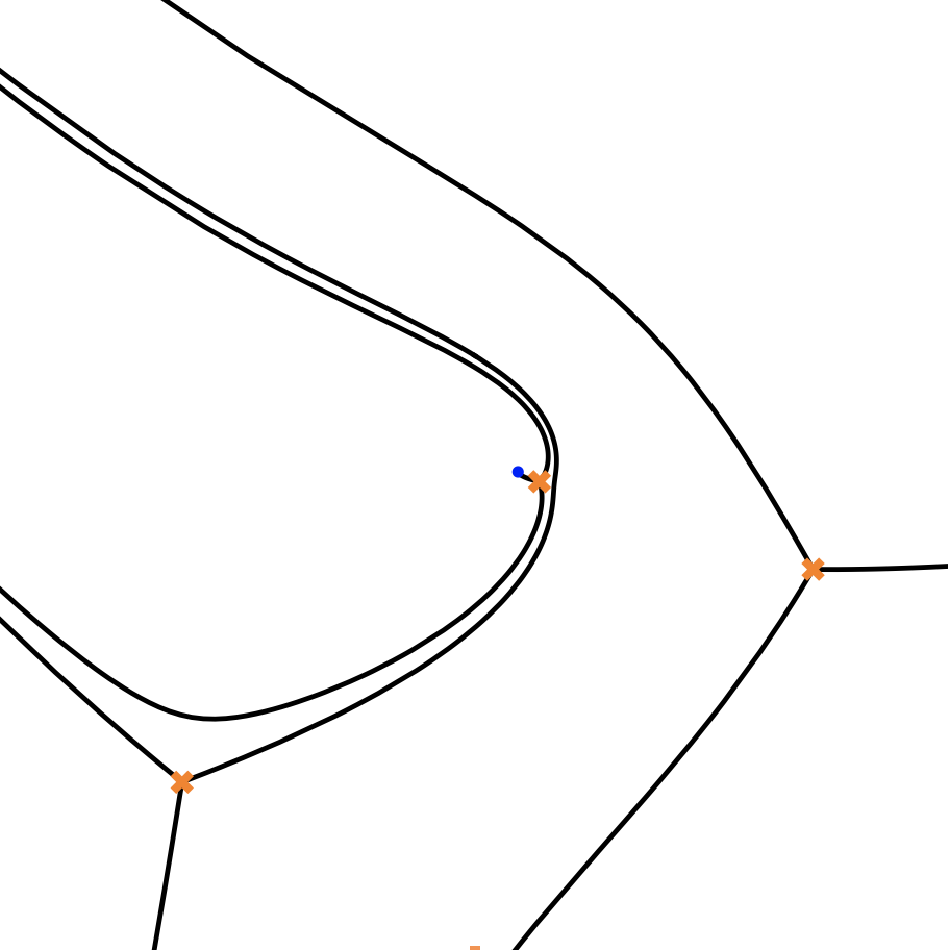}
     \label{fig:fig24}
    \end{minipage}
    }
    \caption{The transition of Stokes graphs for $\xi$-equation with rotating $\theta=\arg(\hbar)$.}
    \label{fig:starkstokes2}
\end{figure}

One may wonder about the continuity of these three EQCs, which is clarified by the saddle reduction of Stokes graphs. It is proven the saddle reduction induced by \eqref{eq: transition} is equivalent to the rotation of $\hbar$. Figure \ref{fig:starkstokes2} shows the transition of Stokes graphs between region $\mathrm{I}_\xi$ and $\mathrm{II}_\xi$ by tuning $F$ and rotating $\theta=\arg(\hbar)$. One starts from a Stokes graph for $F<F_{\xi,1}(1)$ with $\arg(\hbar)=0$ and $m=1$ fixed, whose mutation by rotating $\hbar$ with a small $\delta$ does not change the configuration of the Stokes graph as indicated in figure 3.\ref{fig:fig21} and 3.\ref{fig:fig22}. The Stokes graph moves to figure 3.\ref{fig:fig23} continuously by increasing F with $\arg(\hbar)$ fixed. Then as we continuously change $\arg(\hbar)$ to $0$ with $F$ fixed, the Stokes graph becomes figure 3.\ref{fig:fig24}. During this process, turning points $b$ and $c$ are connected by one Stokes line at an angle $\varphi$ with $0<\varphi<\delta$. And quantum period $\Pi_\gamma(\hbar)$ exhibits singularities in its Borel plane at direction $\varphi$, the discontinuity of the quantum period is given by  the Delabaere-Pham (DP) formula \cite{DDP22}
\begin{equation}
\label{eq:dp}
\exp\left(\frac{1}{\hbar}\mathcal{S}_{+}\left[\Pi_{\gamma}\right]\right)=\exp\left(\frac{1}{\hbar}\mathcal{S}_{-}\left[\Pi_{\gamma}\right]\right)\left(1+\exp\left(-\frac{1}{\hbar}\mathcal{S}\left[\Pi_{\gamma_1}\right]\right)\right),
\end{equation}
here $\mathcal{S}_{\pm}$ are two lateral Borel summations for $\theta=0$ defined in section \ref{sec:wkb}. 
The DP formula assures the EQC or complex resonant frequency is continuous during the transition of the Stokes graphs. Furthermore, it gives the resurgent structure of the quantum periods which indicates the discontinuity of the perturbative contribution is governed by the non-perturbative term. See \cite{Imaizumi,imaizumithesis} for a similar discussion. The continuity of the EQCs between region $\mathrm{II}_\xi$ and $\mathrm{III}_\xi$ can be clarified following the same argument.

The derivation of quantization conditions for $\eta$-equation is parallel to the above analysis.
The boundary condition sorts out the $\psi^{+}(\eta)$ component at origin and infinity, and there is no relevant branch cut to cross. One can write down the relation of the wave functions along the continuation path by matrix multiplication 
\begin{equation}
\begin{aligned}
\left(\begin{array}{l}
\psi_{a, \mathrm{I}_\infty}^{+}(\eta) \\
\psi_{a, \mathrm{I}_\infty}^{-}(\eta)
\end{array}\right)&=M_{+}^{-1} N_{a c} M_{-} N_{c b}M_{+}^{-1}N_{b c}N_{c a}\left(\begin{array}{l}
\psi_{a, \mathrm{I}_0}^{+}(\eta) \\
\psi_{a, \mathrm{I}_0}^{-}(\eta)
\end{array}\right)\\
&=\left(\begin{array}{c}
(1+\mathcal{V}_{\gamma_1}^{-1})\psi_{a, \mathrm{I}_0}^{+}(\eta)-i(1+\mathcal{V}_{\gamma_2}+\mathcal{V}_{\gamma_1}\mathcal{V}_{\gamma_2}) \psi_{a, \mathrm{I}_0}^{-}(\eta) \\
i\mathcal{V}_{\gamma_1}^{-1}\psi_{a, \mathrm{I}_0}^{+}(\eta)+(1+\mathcal{V}_{\gamma_2})\psi_{a, \mathrm{I}_0}^{-}(\eta)
\end{array}\right)
\end{aligned}
\end{equation}
for region $\mathrm{I}_\eta$
\begin{equation}
\left(\begin{array}{l}
\psi_{a, \mathrm{I}_\infty}^{+}(\eta) \\
\psi_{a, \mathrm{I}_\infty}^{-}(\eta)
\end{array}\right)=M_{+}^{-1} N_{a b} M_{+}^{-1} N_{b a}\left(\begin{array}{l}
\psi_{a, \mathrm{I}_0}^{+}(\eta) \\
\psi_{a, \mathrm{I}_0}^{-}(\eta)
\end{array}\right)
=\left(\begin{array}{c}
\psi_{a, \mathrm{I}_0}^{+}(\eta)-i(1+\mathcal{V}_{\gamma}) \psi_{a, \mathrm{I}_0}^{-}(\eta) \\
\psi_{a, \mathrm{I}_0}^{-}(\eta)
\end{array}\right)
\end{equation}
for region $\mathrm{II}_\eta$ and 
\begin{equation}
\begin{aligned}
\left(\begin{array}{l}
\psi_{a, \mathrm{I}_\infty}^{+}(\eta) \\
\psi_{a, \mathrm{I}_\infty}^{-}(\eta)
\end{array}\right)&=M_{+}^{-1} N_{a c} M_{+}^{-1} N_{c b}M_{+}^{-1}N_{b c}N_{c a}\left(\begin{array}{l}
\psi_{a, \mathrm{I}_0}^{+}(\eta) \\
\psi_{a, \mathrm{I}_0}^{-}(\eta)
\end{array}\right)\\
&=\left(\begin{array}{c}
\psi_{a, \mathrm{I}_0}^{+}(\eta)-i(1+\mathcal{V}_{\gamma_2}+\mathcal{V}_{\gamma_1}\mathcal{V}_{\gamma_2}) \psi_{a, \mathrm{I}_0}^{-}(\eta) \\
\psi_{a, \mathrm{I}_0}^{-}(\eta)
\end{array}\right)
\end{aligned}
\end{equation}
for region $\mathrm{III}_\eta$. We use $M_{+}^{-1}$ because of the clockwise continuation. The quantization condition can be written as
\begin{align}
\label{eq:qc0}
&\Pi_\gamma(\hbar)+\log\left(1+e^{-\frac{1}{\hbar}\Pi_{\gamma_{1}}(\hbar)}\right)=2 \pi i \hbar\left(n_\eta+\frac{1}{2}\right),\,\text{for region}\, \mathrm{I}_\eta,\\
\label{eq:qc3}
&\Pi_\gamma(\hbar)=\oint P_{\mathrm{even}}(\eta) d \eta=2 \pi i \hbar\left(n_\eta+\frac{1}{2}\right),\,\text{for region}\, \mathrm{II}_\eta,\\
\label{eq:qc4}
&\Pi_\gamma(\hbar)+\log\left(1+e^{-\frac{1}{\hbar}\Pi_{\gamma_{1}}(\hbar)}\right)=2 \pi i \hbar\left(n_\eta+\frac{1}{2}\right),\,\text{for region}\, \mathrm{III}_\eta,
\end{align}
where $n_\eta$ takes values in non-negative integers. The quantum periods are Borel-resummed as well.
Similarly, the last term contributes exponentially small, it can be neglected in the semi-classical limit. The continuity of the two EQCs comes from \eqref{eq:dp} by a similar argument.
\subsection{Computation and results}
Now we utilize the quantization conditions in the last subsection to compute complex resonant frequencies and compare them with the
results from the Riccati-Pad{\'e} method (RPM) \cite{rpm} that shows very high precision.
We first establish the classical periods in terms of elliptic integrals.
\paragraph{Classical periods} 
Let us parametrize the potential $Q_0(\xi)$ in the $\xi$-equation as
\be
Q_0(\xi)=\frac{u_0\xi^3+u_1\xi^2+u_2 \xi+u_3}{\xi^2}=\frac{u_0(\xi-a)(\xi-b)(\xi-c)}{\xi^2},
\ee
where the coefficients in the potential are related to Stark parameters as \eqref{eq:qlangerxi}.
We consider the relevant period corresponding to one-cycle $\gamma$ which appears in the quantization condition. The integrals for one-cycle $\gamma_1$ can be obtained by the rotation of roots $a\to b$, $b\to c$ and $c\to a$. The classical period is
\be
\ba
\Pi_{\xi}^{(0)}&=2 \sqrt{u_0}\int_b^a \frac{d \xi}{\xi} \sqrt{(\xi-a)(\xi-b)(\xi-c)}\\
&=4 i\sqrt{\frac{u_0}{a-c}}\left(-s \mathbf{K}(k)+t \mathbf{E}(k)+b c \mathbf{\Pi}\left(\alpha^2, k\right)\right),
\ea
\ee
where
\begin{equation}
\begin{aligned}
\label{eq:sub}
s & =\frac{b(c+2 a)+c(a-c)}{3}=\frac{-c^2+a c+b c+2 a b}{3}, \\
t & =-\frac{(c-a)(a+b+c)}{3}, \quad
k^2 =\frac{a-b}{a-c}, \quad
\alpha^2 =\frac{a-b}{a}.
\end{aligned}
\end{equation}
$\mathbf{K}(k)$, $\mathbf{E}(k)$, and $\mathbf{\Pi}\left(\alpha^2, k\right)$ are the elliptic integrals of the first, second, and third kind \cite{gradshteyn2014table}. 
The numerical calculation is a bit subtle since the turning points and integral contours are all in the complex coordinate plane, one should be careful about the branch cuts and sheets. Similarly, for the $\eta$-equation, we parametrize the potential $Q_0(\eta)$ in \eqref{eq:qlangereta}
as
\be
Q_0(\eta)=\frac{u_0\eta^3+u_1\eta^2+u_2 \eta+u_3}{\eta^2}=\frac{u_0(\eta-a)(\eta-b)(\eta-c)}{\eta^2}.
\ee
The classical period is evaluated as
\be
\ba
\Pi_{\eta}^{(0)}&=2  \sqrt{u_0}\int_a^b \frac{d \eta}{\eta} \sqrt{(\eta-a)(\eta-b)(\eta-c)}\\
&=4\sqrt{\frac{u_0}{c-a}}\left\{s \mathbf{K}(k)-t \mathbf{E}(k)-b c \mathbf{\Pi}\left(\alpha^2, k\right)\right\},
\ea
\ee
here $u_0=-\frac{F}{4}$ is negative, which gives an overall imaginary unit $i$. These formulas have already appeared in the literature like \cite{starkwkb82}, but the conventions are different from an imaginary unit $i$. It is almost impossible to compute quantum period by direct integration, yet there is a systematic method based on the Picard-Fuchs equation to evaluate these quantities. It can be sketched by introducing the following fundamental periods.

\paragraph{Fundamental periods} 
We first define the differential 
\be
\ba
\lambda_1&=\partial_{u_1}\sqrt{Q_0(x)} dx=\frac{1}{2\sqrt{Q_0(x)}}dx,\\
\lambda_2&=\partial_{u_2}\sqrt{Q_0(x)} dx=\frac{1}{2x\sqrt{Q_0(x)}}dx,\\
\lambda_3&=\partial_{u_3}\sqrt{Q_0(x)} dx=\frac{1}{2x^2\sqrt{Q_0(x)}}dx.
\ea
\ee
We use $x$ to represent $\xi$ or $\eta$ uniformly for simplicity. It was found that all $p_{2n}(x)$ can be expressed by the combination of the above fundamental differentials up to a total derivative
\begin{equation}
\label{eq:prelation}
    p_{2n}(x)=\sum_{i=1}^{3}c_i^{(n)}\lambda_i+d(*).
\end{equation}
The $\hbar$ expansions for the quantum periods can be evaluated by determining the Picard-Fuchs coefficients $c_i^{(n)}$
and the fundamental periods, which are defined as
\begin{equation}
 (\Pi_\xi)_i=\oint_{\gamma}\lambda_i, \quad i=1,2,3.
\end{equation}
It can be evaluated in terms of elliptic integrals
\be
\ba
(\Pi_\xi)_1&=\oint_\gamma \lambda_1=\int_{b}^{a}\frac{d \xi}{\sqrt{Q_0(\xi)}}=\frac{i}{\sqrt{u_0}}\left(\frac{c}{\sqrt{a-c}}\mathbf{K}(k)+\sqrt{a-c}\mathbf{E}(k)\right), \\
(\Pi_\xi)_2&=\oint_\gamma \lambda_2=\int_{b}^{a}\frac{d \xi}{\xi \sqrt{Q_0(\xi)}}=\frac{i}{\sqrt{u_0}}\frac{1}{\sqrt{a-c}}\mathbf{K}(k), \\
(\Pi_\xi)_3&=\oint_\gamma \lambda_3=\int_{b}^{a}\frac{d \xi}{\xi^2\sqrt{Q_0(\xi)}}=\frac{i}{\sqrt{u_0}}\frac{1}{a\sqrt{a-c}}\mathbf{\Pi}(\alpha^2,k).
\ea
\ee
Similarly, the fundamental periods for $\eta$ coordinate are evaluated as follows
\be
\ba
(\Pi_\eta)_1&=-\frac{1}{\sqrt{u_0}}\left(\frac{c}{\sqrt{c-a}}\mathbf{K}(k)-\sqrt{c-a}\mathbf{E}(k)\right), \\
(\Pi_\eta)_2&=-\frac{1}{\sqrt{u_0}}\frac{1}{\sqrt{c-a}}\mathbf{K}(k), \quad
(\Pi_\eta)_3=-\frac{1}{\sqrt{u_0}}\frac{1}{a\sqrt{c-a}}\mathbf{\Pi}(\alpha^2,k),
\ea
\ee
here $u_0$ is negative as well.
Then the quantum periods are given by 
\begin{equation}
    \Pi_{\xi,\eta}(\hbar)= \sum_{n=0}^{\infty}\hbar^{2n}\left(c_1^{(n)}(\Pi_{\xi,\eta})_1+c_2^{(n)}(\Pi_{\xi,\eta})_2+c_3^{(n)}(\Pi_{\xi,\eta})_3\right),
\end{equation}
where the first several order Picard-Fuchs coefficients are listed as follows
\begin{equation}
\label{eq:pf}
\begin{aligned}
c_1^{(0)}&= \frac{2 u_1}{3},\quad c_2^{(0)}=\frac{4u_2}{3},\quad
c_3^{(0)}=2 u_3 ,\\
  c_1^{(1)}&= \frac{u_0 \left(u_2\left(u_1^2-6 u_0 u_2\right)+9
  u_0 u_1 u_3\right)}{12 \left(-18 u_0u_1u_2
   u_3+u_0\left(27 u_0 u_3^2+4 u_2^3\right)+4
   u_1^3 u_3-u_1^2 u_2^2\right)},\\
   c_2^{(1)}&=-\frac{u_0\left(9
   u_0 u_2 u_3-6 u_1^2 u_3+u_1
 u_2^2\right)}{12 \left(-18 u_0 u_1u_2u_3+u_0
   \left(27 u_0u_3^2+4u_2^3\right)+4 u_1^3
  u_3-u_1^2 u_2^2\right)},\quad
  c_3^{(1)}=0.   
\end{aligned}
\end{equation}
The higher-order coefficients are  too long to list here. We use short notations 
$u_0=\frac{F}{4}$, $u_1=-\frac{E}{2}$, $u_2=-A_1$ and $u_3=\frac{m^2}{4}$ for the $\xi$-equation, $u_0=-\frac{F}{4}$, $u_1=-\frac{E}{2}$, $u_2=-(1-A_1)$ and $u_3=\frac{m^2}{4}$ for the $\eta$-equation here. The computation indicates all $c_3^{(n)}=0$ for $n>0$. All these coefficients are independent of 1-cycle from its definition, so they also apply to $\gamma_1$ and $\gamma_2$.

\paragraph{Numerical calculations}
Although we have derived the EQCs in terms of Borel-resummed quantum periods, it is of course impossible to calculate all-order quantum corrections explicitly. However one can solve the EQCs by accounting for the first several orders of quantum periods. Since the divergence of all-order expansions of quantum periods, numerical techniques such as the Borel-Pad{\'e} method are necessary.

It is straightforward to calculate the complex energy by simultaneously solving quantization conditions for $\xi$- and $\eta$-equations. We show the numerical results for $F=0.03, 0.1, 10$ and $m$ from $0$ to $3$ with half unit in table \ref{table:qc1}. As aforementioned, $\hbar$ in the exact quantization conditions are set to 1 in the numerical calculations. $m=0$ is a special case where the turning point $b$ collapses to $0$, and the elliptic integral of the third kind $\mathbf{\Pi}(\alpha^2,k)$ in $\Pi_{\gamma}(\hbar)$ becomes zero. The authors in \cite{starkwkb82} adopted the BS quantization condition with $b \to 0$ limit, but the numerical test suggests it is better to incorporate an additional term regardless of $F$. Specifically, the EQCs for the $m=0$ case can be written as 
\begin{equation}
\Pi_{\gamma}(\hbar)+\log\left(1+\exp\left(-\frac{1}{\hbar}\Pi_{\gamma_{12}}(\hbar)\right)\right)=2\pi i\hbar\left(n+\frac{1}{2}\right),
\end{equation}
where $\Pi_{\gamma_{12}}(\hbar)$ is the quantum period corresponding to the one-cycle encircling two distinct turning points. It is computed as $\Pi_{\gamma}(\hbar)$ in the next section, we use subscript $\gamma_{12}$ to distinguish it from $\gamma$ in this section.

\begin{table}[htbp]
\centering
\begin{tabular}{c c c c }
\hline
$F$&  0.03 & 0.1 & 10\\
\hline
$m=0$& $-0.500961 - 2.151[8]i$ & $-0.517428 -0.016599 i$ & $0.576449 - 4.28595 i$ \\
 &$-0.502074 - 1.119[8] i$ & $-0.527418- 0.007269 i$ & $0.608272 - 5.57802 i $\\
 \hline
$m=\frac{1}{2}$
 &$-0.226066-0.047773 i$ & $-0.274726-0.092727 i$ & $2.03209-7.57066 i$ \\
 
 &$-0.240498-0.007357 i$ & $-0.275297-0.099408 i$&  $2.29507-7.89113 i$ \\
 \hline
 $m=1$
 &$-0.153377-0.041476 i$ & $-0.157815-0.197065 i$ & $3.37685-9.53549 i$ \\
 
 &$-0.153357-0.044424 i$ & $-0.154012-0.201879 i$ & $3.59267-9.82232 i$ \\
 \hline
 $m=\frac{3}{2}$ 
 &$-0.103499-0.084042 i$ & $-0.071613-0.289995 i$ & $4.50685-11.2655 i$ \\
 &$-0.101927-0.085442 i$ & $-0.070198-0.293556 i$ & $4.69979-11.5350 i $\\
 \hline
 $m=2$
 &$-0.065022-0.122749 i$ & $-0.003934-0.373115 i$ & $5.51089-12.8427 i $\\

 &$-0.064236-0.123769 i$ & $-0.003500-0.376159 i$ & $5.68990-13.1015 i$ \\
 \hline
$m=\frac{5}{2}$
 &$-0.034074-0.158318 i$ & $0.053511-0.449190 i$ & $6.42954-14.3096 i$ \\
 &$-0.033645-0.159169 i$ & $0.053447-0.451947 i$ & $6.59891-14.5609 i$ \\
 \hline
$m=3$
 &$-0.007590-0.191361 i$ & $0.104412-0.519933 i$ & $7.28523-15.8465 i $\\
 &$-0.007353-0.192112 i$ & $0.104058-0.522512 i$ & $7.44738-15.9374 i$ \\
 \hline
\end{tabular}
\caption{The table shows complex resonant frequencies $E$ for distinct magnetic quantum number $m$ with electric field strength from weak to very strong. The first row stands for the results computed from quantization conditions \eqref{eq:qc1} and \eqref{eq:qc3} for $F=0.03, 0.1$, and conditions \eqref{eq:qc2} and \eqref{eq:qc4} for $F=10$. $m=0$ case is calculated separately as explained in the main content. Here only classical periods $\Pi^{(0)}$ are considered. The second row is for the Riccati-Pad{\'e} method. [a] means $10^{-a}$, this notation is also used later.}
\label{table:qc1}
\end{table}

As for nonzero $m$, we adopt exact quantization conditions for proper regions derived in the previous section. We note that the two equations may take different types of quantization conditions even for the same $F$ and $m$ from figure \ref{fig: transition}. At this stage, we only consider the first-order contribution, namely $\Pi^{(0)}$. For convenience to compare with other methods, both quantum numbers $n_\xi$ and $n_\eta$ are set to be $0$. It is the worst case for applying the WKB method since the WKB method approximates much better for large $n$ states, whose calculations are also straightforward. We find even the first-order WKB analysis gives good estimation for a wide range of parameter choices. $F=0.03$ is already a strong field intensity in the atomic unit. If one examines for very weak field $F$, the result shows high precision as in \cite{starkwkb82}. The calculation indicates even for a very strong field, for example, $F=10$ here, the approximation is also efficient.

We also find the effect of the logarithmic term in modified BS quantization conditions. Table \ref{table:qc2} shows the complex energies for a strong field intensity $F=5$ and $m=1,2,3$. Quantization conditions including the logarithmic term give closer results to exact values. Our observation further implies the imaginary part gets corrected significantly by the additional modified term.

\begin{table}[H]
\centering
\begin{tabular}{c c c c}
\hline
$F=5$&  $m=1$ & $m=2$&$m=3$\\
\hline

1st-order BSQC
 &$2.32739-7.03055 i$ & $3.54168-9.13333 i$ & $4.58412-10.9252 i$ \\
 
1st-order EQC
 &$1.70365-5.72785 i$ & $3.12615-7.76650 i$ & $4.26965-9.58103 i$ \\
 
 RPM& $1.87776-5.82185 i$ & $3.24425-7.91809 i$ & $4.37684-9.72372 i$ \\
 \hline
\end{tabular}
\caption{This shows the complex resonant frequencies for $F=5$ and $m=1,2$ and $3$. The first and second row represents the results from BS quantization conditions and exact quantization conditions respectively. Here only contributions from classical periods $\Pi^{(0)}$ are included.}
\label{table:qc2}
\end{table}

One of the new results in this paper is to include the quantum corrections to $\Pi^{(0)}$. Table \ref{table:qc3} exhibits the results for $F=0.005$ by including the first several orders of quantum periods. The energy level for the ground state is really precise, and the imaginary part of complex energy for this case is very small\footnote{approximately $10^{-56}$ as shown in \cite{rpm2}} and neglected here. However, the improvement for $m=3$ is not so satisfactory. One quick reason may be that the quantum corrections are very large compared with classical periods for this case, where the numerical approximation does not perform very well even when the Borel-Padé technique is adopted.

\begin{table}[htp]
    \centering
    \begin{tabular}{c c c}
    \hline
     magnetic quantum number($m$)   & $0$ & $3$  \\
     \hline
       $\mathcal{O}(\hbar^0)$  & $-0.5000265658500433$  & $-0.0402674 - 0.0326842 i$ \\
  
     $\mathcal{O}(\hbar^2)$  & $-0.5000562669655335$  & $-0.0401851 - 0.0327634 i$ \\

      $\mathcal{O}(\hbar^4)$   & $-0.5000562847698887$ & $-0.0401867 - 0.0327546 i$ \\
    
     $\mathcal{O}(\hbar^6)$ & $-0.5000562847937314$ &  $-0.0401877-0.0327561i$\\
    
 $\mathcal{O}(\hbar^8)$  & $-0.5000562847937927$ &$-0.0401869-0.0327561 i$ \\

     RPM & $-0.5000562847937930$  &  $-0.0400286 - 0.0328640 i$ \\
    \hline
    \end{tabular}
    \caption{Complex energy frequencies for electric field strength $F=0.005$ with magnetic quantum number $m=0$ and $m=3$ respectively. Here we use BS quantization conditions for both cases since the logarithmic corrections for $m=0$ are very tiny and ignorable.}
    \label{table:qc3}
\end{table}

It is not expected in general that the present method entirely matches with standard perturbation theory results for each order of $\hbar$ except for the first order. Since our method is based on Langer's modification which is just a kind of expansion scheme. Different schemes give different results, but they all coincide to include all-order $\hbar$ expansions. In practice, we use the first several leading orders to approximate the exact Borel-resummed quantum WKB periods as a numerical approach.

\section{TBA equations and analytic continuation}
\label{sc:tba}
The quantum period is the fundamental ingredient in the WKB analysis. However, it can also be formulated into integral equations, namely, TBA equations from an alternative approach based on the ODE/IM correspondence. The solutions of the TBA equations reproduce the Borel-resummed quantum periods exactly. TBA equations including centrifugal terms have been discussed in \cite{Ito-Shu} and \cite{xiyin}. They correspond to the Stark problem without Langer's modification, which implies the different (unphysical) boundary conditions at the origin for the wave function and different WKB cycles. These TBA equations are therefore  not relevant to the spectral problem of the Stark effect with Langer's modification for general $m$. However, for $m=0$, both setups are the same. Then we can check the WKB correction of quantum periods precisely.

\subsection{Singular potentials and quantum periods}

Let us now revisit the Stark problem without Langer's modification to reconsider \eqref{eq:parabolicseq}, which is uniformly written as \eqref{eq:potential}. We will consider quantum periods from the WKB expansions at first. We use the same notations for $Q_{0}(x), 
\Pi(\hbar)$, and $\gamma$ which should not be confused with the prescription adopted in the last section. 
The classical and quantum periods can be written in terms of elliptic integrals.
\paragraph{Classical periods} Denote two turning points $e_1$ and $e_2$ of $Q_0(x)$ in \eqref{eq:potential}:
\be
\label{eq:tp}
Q_0(x)=\frac{u_0x^2+u_1x+u_2}{x}=\frac{u_0(x-e_1)(x-e_2)}{x},\quad e_{1,2}=\frac{-u_1\pm\sqrt{u_1^2-4u_0u_2}}{2u_0}.
\ee
One can define three one-cycles on the Riemann surface defined by $y^2=Q_0(x)$, which encircle $e_1$ and $e_2$, $0$ and $e_1$, $0$ and $e_2$ and are denoted as $\gamma$, $\hat{\gamma}_1$ and $\hat{\gamma}_2$ respectively. Then the classical periods corresponding to these three cycles are represented by the elliptic integrals as
\be
\ba
\Pi_{\gamma}^{(0)}&=2\int_{e_1}^{e_2}\sqrt{Q_0(x)}d x
=2i (e_2-e_1)^2\sqrt{\frac{u_0}{e_1}}\frac{4(2+k)\mathbf{E}(-k)-8(1+k)\mathbf{K}(-k)}{6k^2},\\
\Pi_{\hat{\gamma}_1}^{(0)}&=2\int_{0}^{e_1}\sqrt{Q_0(x)}d x=\frac{4}{3}\sqrt{u_0 e_1^{3} m}\left((1+m)\mathbf{E}(\frac{1}{m})-(m-1)\mathbf{K}(\frac{1}{m})\right),\\
\Pi_{\hat{\gamma}_2}^{(0)}&=2\int_{e_2}^{0}\sqrt{Q_0(x)}d x=-\frac{4}{3}\sqrt{\frac{u_0 e_2^{3}}{m} }\left((1+\frac{1}{m})\mathbf{E}(m)-(\frac{1}{m}-1)\mathbf{K}(m)\right),
\ea
\ee
where $m=\frac{e_2}{e_1}$ and $k=\frac{e_2}{e_1}-1$. These periods are related by their definitions 
\begin{equation}
\Pi_{\gamma}^{(0)}+\Pi_{\hat{\gamma}_1}^{(0)}+\Pi_{\hat{\gamma}_2}^{(0)}=0.
\end{equation}
The quantum periods are evaluated by the Picard-Fuchs method as well, which is composed of a combination of fundamental periods. 
\paragraph{Fundamental periods} Define the fundamental differentials first for $Q_0(x)$:
\be
\ba
\lambda_1&=\partial_{u_1}\sqrt{Q_0(x)} dx=\frac{1}{2\sqrt{Q_0(x)}}dx,\\
\lambda_2&=\partial_{u_2}\sqrt{Q_0(x)} dx=\frac{1}{2x\sqrt{Q_0(x)}}dx,\\
\ea
\ee
and the associated fundamental periods
\begin{equation}
    (\Pi_\gamma)_i=\oint_{\gamma}\lambda_i.
\end{equation}
They are represented as 
\be
(\Pi_\gamma)_1=\int_{e_1}^{e_2}\frac{d x}{\sqrt{Q_0(x)}}=-2i\sqrt{\frac{e_1}{u_0}}\mathbf{E}(-k),\quad
(\Pi_\gamma)_2=\int_{e_1}^{e_2}\frac{d x}{x\sqrt{ Q_0(x)}}=-\frac{2i}{\sqrt{u_0 e_1}} \mathbf{K}(-k),
\ee

\begin{equation}
(\Pi_{\hat{\gamma}_1})_1=\int_{0}^{e_1}\frac{d x}{\sqrt{Q_0(x)}}=2\sqrt{\frac{u_0 m e_1^3}{u_1^2}}\left((1+m)\mathbf{E}(\frac{1}{m})-(m-1)\mathbf{K}(\frac{1}{m})\right)-\frac{u_2}{u_1}\frac{4}{\sqrt{u_0 e_1 m}}\mathbf{K}(\frac{1}{m}),
\end{equation}

\begin{equation}
  (\Pi_{\hat{\gamma}_1})_2=\int_{0}^{e_1}\frac{d x}{x\sqrt{ Q_0(x)}}=\frac{2}{\sqrt{u_0 e_1 m}}\mathbf{K}(\frac{1}{m}),
\end{equation}

\begin{equation}
  (\Pi_{\hat{\gamma}_2})_1=\int_{e_2}^{0}\frac{d x}{\sqrt{ Q_0(x)}}=\frac{4u_2}{u_1}\sqrt{\frac{m}{u_0 e_2}}\mathbf{K}(m)-2\sqrt{\frac{u_0 e_2^3}{m u_1^2}}\left((1+\frac{1}{m})\mathbf{E}(m)-(\frac{1}{m}-1)\mathbf{K}(m)\right),
\end{equation}

\begin{equation}
  (\Pi_{\hat{\gamma}_2})_2=\int_{e_2}^{0}\frac{d x}{x\sqrt{ Q_0(x)}}=-2\sqrt{\frac{m}{u_0 e_2}}\mathbf{K}(m).
\end{equation}
As the same as \eqref{eq:prelation}, there is a similar relation
\begin{equation}
    p_{2n}(x)=\sum_{i=1}^{2}c_i^{(n)}\lambda_i+d(*),
\end{equation}
which suggests writing down quantum periods by
\begin{equation}
    \Pi(\hbar)= \sum_{n=0}^{\infty}\hbar^{2n}\left(c_1^{(n)}(\Pi)_1+c_2^{(n)}(\Pi)_2\right),
\end{equation}
where the subscript $\gamma, \hat{\gamma}_1$, and $\hat{\gamma}_2$ are ommited. The Picard-Fuchs coefficients $c_{1,2}^{(n)}$
can be obtained similarly as \eqref{eq:pf}, they are of course different although we abuse the same notations. The first several coefficients are listed as follows
\begin{equation}
    \begin{aligned}
        c_1^{(0)}&=\frac{2 u_1}{3},\quad  c_2^{(0)}=\frac{4 u_2}{3},\\
        c_1^{(1)}&= \frac{1}{6} u_0 \left(\frac{6\ell(\ell+1)+1}{u_2}+\frac{u_0}{u_1^2-4 u_0
   u_2}\right),\quad
        c_2^{(1)}=\frac{u_0 u_1}{12 \left(u_1^2-4 u_0 u_2\right)},
    \end{aligned}
\end{equation}
where $\ell$ is related to $m$ by \eqref{eq:lm}.
\subsection{TBA equations for the minimal chamber}
Now we study the quantum period with its relation to the TBA equations. Let us consider
\begin{equation}
\label{eq:sseq}
\left(-\hbar^{2} \frac{d^2}{dx^2}+u_0 x+u_1+\frac{u_2}{x}+\frac{\hbar^{2} \ell(\ell+1)}{x^{2}}\right) \psi(x)=0.
\end{equation}
The TBA equations for \eqref{eq:sseq} within $-1<\ell<1$ were first proposed in \cite{Ito-Shu}.
\begin{equation}
\label{eq:tba1}
\begin{aligned}
& \log Y_1(\theta)=-m_1 e^\theta+\int_{-\infty}^{\infty} \frac{d \theta^{\prime}}{2 \pi} \frac{\log \left(1-e^{2 \pi i \ell} \hat{Y}\left(\theta^{\prime}\right)\right)\left(1-e^{-2 \pi i \ell} \hat{Y}\left(\theta^{\prime}\right)\right)}{\cosh \left(\theta-\theta^{\prime}\right)}, \\
& \log \hat{Y}(\theta)=-\hat{m} e^\theta+\int_{-\infty}^{\infty} \frac{d \theta^{\prime}}{2 \pi} \frac{\log \left(1+Y_1(\theta^\prime)\right)}{\cosh \left(\theta-\theta^{\prime}\right)}.
\end{aligned}
\end{equation}
Here $-\log Y_1(\theta)$ and $-\log \hat{Y}(\theta)$ correspond to the energies of the pseudo-particles in the integrable system.
A new parameter $\theta=-\log\hbar$ called the rapidity is introduced, which takes the semi-classical limit $\hbar\to0$ to $\theta\to\infty$. The leading contribution in the large $\theta$ expansion of the $Y$ functions is captured by the classical period
\begin{equation}
\begin{aligned}
&Y_1\left(\theta\right) \sim e^{-\frac{m_1}{\hbar}}, \quad m_1 \equiv -i \oint_{\gamma} \sqrt{Q_0(x)} d x=-i\Pi_{\gamma}^{(0)},\\
    &\hat{Y}\left(\theta\right) \sim e^{-\frac{\hat{m}}{\hbar}}, \quad \hat{m} \equiv  \oint_{\hat{\gamma}_1} \sqrt{Q_0(x)} d x=\Pi_{\hat{\gamma}_1}^{(0)}.
\end{aligned}
\end{equation}
Moreover, the higher-order expansions of the $Y$-function for large $\theta$ are consistent with all-order quantum periods in the viewpoint of their analyticity and asymptotics. When the mass parameters $m_1$ and $\hat{m}$ are complex, we set
\begin{equation}
m_1=\left|m_1\right| e^{i \phi_1}, \quad \hat{m}=|\hat{m}| e^{i \hat{\phi}},
\end{equation}
then the driving term in \eqref{eq:tba1} becomes $m_1 e^\theta=|m_1|e^{\theta+i\phi_1}$ and $\hat{m}e^{\theta}=|\hat{m}|e^{\theta+i\hat{\phi}}.$
We shift $\theta \to \theta-i\phi_1$ for $Y_1(\theta)$ and $\theta \to \theta-i\hat{\phi}$ for $\hat{Y}(\theta)$ so that the TBA equations becomes
\begin{equation}
\label{eq:tba2}
\begin{aligned}
\log Y_1\left(\theta-i \phi_1\right) & =-\left|m_1\right| e^\theta+\int_{-\infty}^{\infty} \frac{d \theta^{\prime}}{2 \pi} \frac{\log \left(1-e^{2 \pi i \ell} \hat{Y}\left(\theta^{\prime}-i\hat{\phi}\right)\right)\left(1-e^{-2 \pi i \ell} \hat{Y}\left(\theta^{\prime}-i\hat{\phi}\right)\right)}{\cosh \left(\theta-\theta^{\prime}-i \phi_1+i \hat{\phi}\right)}, \\
\log \hat{Y}(\theta-i \hat{\phi}) & =-|\hat{m}| e^\theta+\int_{-\infty}^{\infty} \frac{d \theta^{\prime}}{2 \pi} \frac{\log \left(1+Y_1\left(\theta-i \phi_1\right)\right)}{\cosh \left(\theta-\theta^{\prime}-i \hat{\phi}+i \phi_1\right)}.
\end{aligned}
\end{equation}
These equations hold when 
\begin{equation}
\label{eq:phasecondition}
    |\phi_1-\hat{\phi}|<\frac{\pi}{2},
\end{equation}
which defines a region in the space of parameters $u_0$, $u_1$,  and $u_2$. It is called the minimal chamber. If we rotate $\hbar$ to redefine a new constant $\Tilde{\hbar}=i\hbar$, then the equation for $\eta$ coordinate becomes
\begin{equation}
\label{eq:singularseq2}
\left(-\Tilde{\hbar}^{2} \frac{\mathrm{d}^{2}}{\mathrm{d} \eta^{2}}+\frac{F}{4}\eta+\frac{E}{2}+\frac{A_{2}}{\eta}+\frac{\Tilde{\hbar}^{2} (m^{2}-1)}{4 \eta^{2}}\right) \psi_2(\eta)=0.
\end{equation}
We found the parameters for the ground state with $m=0$ lie in the minimal chamber now.
We solve these TBA equations  iteratively via the Fourier discretization by taking $m_1$ and $\hat{m}$ as the initial seed and compute the first several order expansions by the large $\theta$ expansion for $\log Y(\theta)$
\begin{equation}
\ba
\label{eq:tbaexpansion}
-\log Y_1(\theta)& \sim m_1 e^\theta+\sum_{n \geq 1} m_1^{(n)} e^{(1-2 n) \theta},\\
-\log \hat{Y}(\theta)& \sim \hat{m}e^\theta+\sum_{n \geq 1} \hat{m}^{(n)} e^{(1-2 n) \theta}.
\ea
\end{equation}
The expansion coefficients are 
\begin{equation}
\begin{aligned}
m_{1}^{(n)} &=\frac{(-1)^{n}}{\pi} \int_{\mathbb{R}} e^{(2 n-1) \theta^\prime}\left[\log \left(1- e^{2 \pi i l} \hat{Y}\left(\theta^{\prime}\right)\right)+\log \left(1-e^{-2 \pi i l} \hat{Y}\left(\theta^{\prime}\right)\right)\right] d \theta^{\prime},\\
\hat{m}^{(n)} &=\frac{(-1)^{n}}{\pi} \int_{\mathbb{R}} e^{(2 n-1) \theta^\prime} \log \left(1+Y_{1}\left(\theta^{\prime}\right)\right) d \theta^{\prime},
\end{aligned}
\end{equation}
here we suppose $m_1$ and $\hat{m}$ are real. The ODE/IM correspondence implies the following relationship
\begin{equation}
\label{eq:correspondence}
   i m_1^{(n)}= (-1)^n\Pi_{\gamma}^{(n)}, \quad \hat{m}^{(n)}=\Pi_{\hat{\gamma}_1}^{(n)},
\end{equation}
where $m_1^{(0)}=m_1$ and $\hat{m}^{(0)}=\hat{m}$ are mass parameters. Numerical results are listed in table \ref{table:qp1} for $\ell=-\frac{1}{2}$ or $m=0$ and $F=0.005$, where $E$ and $A_1$ are substituted as input parameters. We utilize the precise value of $E$ in table \ref{table:qc1} and $A_1$ from the Riccati-Padé method to compute the classical period $\Pi^{(0)}$, which equivalently gives mass parameters $m$ and $\hat{m}$. Large $\theta$ expansions $m^{(n)}$ are compared with the quantum periods $\Pi^{(n)}$ calculated by the Picard-Fuchs method for the first several orders to find them in good agreement.
\begin{table}[h]
    \centering
    \begin{tabular}{c c c c c}
    \hline 
    $n$ & $m_1^{(n)}$ & $\hat{m}^{(n)}$  & $\Pi_{\gamma}^{(n)}$ & $\Pi_{\hat{\gamma}_1}^{(n)}$\\
            \hline 
            $1$ &   $-0.17229336876340$  &  $-0.00405302352020$  & $0.17229336876307 i$& $-0.00405302352018$ \\
            $2$ &  $0.03925568525637$ & $6.009535667649[7]$ &$0.03925568525630 i$& $6.009535667620[7]$ \\
             $3$ &  $-0.04967657873047$ & $-4.680585382632[10]$ & $0.04967657873075 i$  & $-4.680585382628[10]$\\
            \hline 
            \end{tabular}
             \caption{The first three order expansions of the quantum periods for $\eta$-equation from TBA equations and the Picard-Fuchs method for $m=0$ and $F=0.005$.}
              \label{table:qp1}
    \end{table}

\paragraph{Continuation of $\ell$}
The authors in \cite{xiyin} found
as $\ell$ goes from $-1<\ell<0$ to $\ell>0$, TBA equations get modified from the singularities of the $Y$ equations in the complex $\theta$ plane \cite{DT3}. When $0< \ell \lesssim 0.28$, which is named the first modified region in \cite{xiyin}, the TBA equations are 
\begin{equation}
\label{eq:tbacorrect1}
\begin{aligned}
\log Y_{1}(\theta) &=-m_{1} e^{\theta}+\int_{\mathbb{R}} \frac{d \theta^{\prime}}{2 \pi} \frac{\log \left(\left(1-e^{2 \pi i \ell} \hat{Y}\left(\theta^{\prime}\right)\right)\left(1-e^{-2 \pi i \ell} \hat{Y}\left(\theta^{\prime}\right)\right)\right)}{\cosh \left(\theta-\theta^{\prime}\right)}+\log \left(\frac{e^{\theta}+i e^{\alpha_{1}(\ell)}}{e^{\theta}-i e^{\alpha_{1}(\ell)}}\right), \\
\log\hat{Y}(\theta) &=-\hat{m} e^{\theta}+\int_{\mathbb{R}} \frac{d \theta^{\prime}}{2 \pi} \frac{\log \left(1+Y_{1}\left(\theta^{\prime}\right)\right)}{\cosh \left(\theta-\theta^{\prime}\right)},
\end{aligned}
\end{equation}
where $\alpha_{1}(\ell)$ is the zero of $1-e^{2 \pi i \ell} \hat{Y}(\theta)$ in the $|\mathrm{Im}(\theta)|<\frac{\pi}{2}$ strip, and is determined by
\begin{equation}
\label{eq:alpha1}
2 \pi i \ell=-\log \hat{Y}\left(\alpha_{1}(\ell)\right)=\hat{m} e^{\alpha_{1}(\ell)}-\int_{\mathbb{R}} \frac{d \theta^\prime}{2 \pi} \frac{\log \left(1+Y_{1}\left(\theta^{\prime}\right)\right)}{\cosh \left(\alpha_{1}(\ell)-\theta^\prime\right)}.
\end{equation}
The algorithm for solving these TBA equations refers to \cite{xiyin}.
In this case, expansions of the quantum periods are also modified by $\alpha_{1}(\ell)$.
\begin{equation}
\begin{aligned}
m_{1}^{(n)} &=(-1)^{n} \left(\frac{1}{\pi}\int_{\mathbb{R}} e^{(2 n-1) \theta^\prime}\left\{\log \left(1- e^{2 \pi i l} \hat{Y}\left(\theta^{\prime}\right)\right)+\log \left(1-e^{-2 \pi i l} \hat{Y}\left(\theta^{\prime}\right)\right)\right\} d \theta^{\prime}+2 i \frac{e^{(2n-1)\alpha_{1}(\ell)}}{2n-1}\right),\\
\hat{m}^{(n)} &=\frac{(-1)^{n}}{\pi} \int_{\mathbb{R}} e^{(2 n-1) \theta^\prime} \log \left(1+Y_{1}\left(\theta^{\prime}\right)\right) d \theta^{\prime}.
\end{aligned}
\end{equation}
However, it seems the parameters $u_0, u_1$, and $u_2$ for $\ell>0$ are already outside the minimal chamber. But we can test the relationship \eqref{eq:correspondence} for some fictitious value, $u_0=1, u_1=-3, u_2=1$ and $\ell=\frac{1}{5}$ to confirm this correspondence holds on with $\ell$ correction. This is not covered in \cite{xiyin}. We list the numeric results in table \ref{table:singularqp1} to compare with those evaluated from the Picard-Fuchs method. See also \cite{yang} for further discussions.

As $\ell$ increases steadily, the second singular point for the $Y$ function gets involved, and the $\ell$ enters into the second modified region $0.28\lesssim \ell \lesssim 0.5$.
The TBA equations for this region are
\begin{equation}
\begin{aligned}
\log Y_{1}(\theta) &=-m_{1} e^{\theta}+\int_{\mathbb{R}} \frac{d \theta^{\prime}}{2 \pi} \frac{\log \left(\left(1-e^{2 \pi i \ell} \hat{Y}\left(\theta^{\prime}\right)\right)\left(1-e^{-2 \pi i \ell} \hat{Y}\left(\theta^{\prime}\right)\right)\right)}{\cosh \left(\theta-\theta^{\prime}\right)}+\log \left(\frac{e^{\theta}+i e^{\alpha_{1}(\ell)}}{e^{\theta}-i e^{\alpha_{1}(\ell)}}\right), \\
\log \hat{Y}(\theta) &=-\hat{m} e^{\theta}+\int_{\mathbb{R}} \frac{d \theta^{\prime}}{2 \pi} \frac{\log \left(1+Y_{1}\left(\theta^{\prime}\right)\right)}{\cosh \left(\theta-\theta^{\prime}\right)}+\log \left(\frac{e^{\theta}+i e^{\hat{\alpha}(\ell)}}{e^{\theta}-i e^{\hat{\alpha}(\ell)}}\right),
\end{aligned}
\end{equation}
where $\alpha_{1}(\ell)$ and $\hat{\alpha}(\ell)$ are determined by
\begin{equation}
\begin{aligned}
&2 \pi i \ell=\hat{m} e^{\alpha_{1}(\ell)}-\int_{\mathbb{R}} \frac{\log \left(1+Y_{1}\left(\theta^{\prime}\right)\right)}{\cosh \left(\alpha_{1}(\ell)-\theta^{\prime}\right)} \frac{d \theta^{\prime}}{2 \pi}-\log \left(\frac{e^{\alpha_{1}(\ell)}+i e^{\hat{\alpha}(\ell)}}{e^{\alpha_{1}(\ell)}-i e^{\hat{\alpha}(\ell)}}\right) \\
&-\pi i=m_{1} e^{\hat{\alpha}(\ell)}-\int_{\mathbb{R}} \frac{\log \left(\left(1-e^{2 \pi i \ell} \hat{Y}\left(\theta^{\prime}\right)\right)\left(1-e^{-2 \pi i \ell} \hat{Y}\left(\theta^{\prime}\right)\right)\right)}{\cosh \left(\hat{\alpha}(\ell)-\theta^{\prime}\right)} \frac{d \theta^{\prime}}{2 \pi}-\log \left(\frac{e^{\hat{\alpha}(\ell)}+i e^{\alpha_{1}(\ell)}}{e^{\hat{\alpha}(\ell)}-i e^{\alpha_{1}(\ell)}}\right) .
\end{aligned}
\end{equation}
In this case, quantum periods are modified by $\alpha_{1}(\ell)$ and $\hat{\alpha}(\ell)$ as follows
\begin{equation}
\begin{aligned}
m_{1}^{(n)} &=\frac{(-1)^{n}}{\pi} \left(\int_{\mathbb{R}} e^{(2 n-1) \theta^\prime}\left\{\log \left(1- e^{2 \pi i l} \hat{Y}\left(\theta^{\prime}\right)\right)+\log \left(1-e^{-2 \pi i l} \hat{Y}\left(\theta^{\prime}\right)\right)\right\} d \theta^{\prime}+2 i \frac{e^{(2n-1)\alpha_{1}(\ell)}}{2n-1}\right)\\
\hat{m}^{(n)} &=\frac{(-1)^{n}}{\pi} \left(\int_{\mathbb{R}} e^{(2 n-1) \theta^\prime} \log \left(1+Y_{1}\left(\theta^{\prime}\right)\right) d \theta^{\prime}+2 i \frac{e^{(2n-1)\hat{\alpha}(\ell)}}{2n-1}\right)
\end{aligned}
\end{equation}
\begin{table}[htp]
            \centering
            \begin{tabular}{c c c c c}
            \hline $n$& $m_1^{(n)}$ & $\hat{m}^{(n)}$ & $\Pi_{\gamma}^{(n)}$ & $\Pi_{\hat{\gamma}_1}^{(n)}$ \\
            \hline 
         $1$ &   $1.4753093535348$ &  $0.0720517742848$ &$-1.4753093522562i$ &  $0.0720517728119$  \\
            $2$ &  $0.2019237645206$  &$0.0000700313547$  &$0.2019237643494i$& $0.0000700315059$\\
            $3$ & $0.2508765451711 $  &  $-0.0045008552377$ &$-0.2508765449614i$&$-0.0045008554033$\\
            $4$ &  $-1.4745761911756$ & $0.0180119220334$ &$-1.4745761905853i$   & $0.0180119208494$\\ 
            \hline 
            \end{tabular}
              \caption{The first four order expansions of the quantum periods for fictitious parameters $u_0=1, u_1=-3, u_2=1$ and $\ell=\frac{1}{5}$.}
        \label{table:singularqp1}
    \end{table}
Table \ref{table:singularqp2} shows the first four expansions of the quantum periods computed from the above expansions and the Picard-Fuchs method. We also adopt unphysical moduli parameters $u_0=1, u_1=-3, u_2=1$, and $\ell=\frac{2}{5}$. They are all consistent for the  given precision.

One can find that the quantum periods calculated directly from the Picard-Fuchs method agree well with the evaluation from the TBA side for wide parameter choices, which verifies the generalized ODE/IM correspondence.
\begin{table}[H]
            \centering
            \begin{tabular}{c c c c c}
            \hline $n$& $m_1^{(n)}$ & $\hat{m}^{(n)}$ & $\Pi_{\gamma}^{(n)}$ & $\Pi_{\hat{\gamma}_1}^{(n)}$ \\
            \hline 
            $1$ &  $2.6562188814454$  & $0.1978678217639$ &$-2.6562188791176i$&$0.19786781876613$\\
            $2$ & $1.9329727067243$  &  $-0.0054720259650$  &$1.9329727050823i$&  $-0.00547202448659$ \\
            $3$ & $1.6659668485245$  &  $0.0005791023097$ &$-1.6659668471070i$& $0.00057910105643$  \\
            $4$ & $6.7514239599655$  & $0.0064465654840$ &$6.7514239558741i$&$0.00644657053159$\\
            \hline 
            \end{tabular}
             \caption{The first four order expansions of the quantum periods for $u_0=1, u_1=-3, u_2=1$ and $\ell=\frac{2}{5}$.}
        \label{table:singularqp2}
    \end{table}

\subsection{Analytic continuation and wall-crossing}
The TBA equations \eqref{eq:tba2} are valid only when \eqref{eq:phasecondition} is satisfied. 
The kernel function 
\begin{equation}
    \frac{1}{\cosh\left(\theta-\theta^\prime-i\phi_1+i\hat{\phi}\right)}
\end{equation}
has poles at
\begin{equation}
\theta^{\prime}=\theta-i \phi_1+i \hat{\phi}+\frac{i \pi}{2}+i n \pi, \quad n \in \mathbb{Z},
\end{equation}
where $\theta-i\phi_1+i\hat{\phi}-\frac{i\pi}{2}$ locates below the real axis and $\theta-i\phi_1+i\hat{\phi}+\frac{i\pi}{2}$ above the real axis. These poles correspond to the singularities of the Borel transforms of the quantum periods. When $\phi_1-\hat{\phi}$ increases and cross over $\phi_1-\hat{\phi}>\frac{\pi}{2}$, the pole $\theta-i\phi_1+i\hat{\phi}+\frac{i\pi}{2}$ moves down and cross the real axis. The $\theta^\prime$-integral picks the residue at this pole. Similarly the second equation in \eqref{eq:tba2} picks the pole at $\theta-i\phi_1+i\hat{\phi}-\frac{i\pi}{2}$. This phenomenon is the wall-crossing of the TBA equations. The TBA equations after wall-crossing are modified as Appendix B in \cite{Ito-Shu}.

\begin{table}[h]
        \centering
        \begin{tabular}{c c c}
         \hline 
         $n$ & $m_1^{(n)}$ & $\hat{m}^{(n)}$\\
          \hline 
         $0$ & $4.3446402338816 + 15.4257466346279i $&  $- 0.0012584924834 -9.4789003437673 i$\\
        $1$ &  $0.0266639256955 - 0.0235576487819i$ &    $0.0012417895471 + 0.0537580797986i$ \\
         $2$ & $-1.62952996522[4]-4.09974572924[4]i$  &$1.7792731990[5] + 3.62538591473[4]i $\\
         \hline
         
         $n$&   $\Pi_{s,1}^{(n)}$&$2\hat{\Pi}_{s}^{(n)}$\\
         \hline
         $0$& $4.3446402284715i-15.4257465764022$&$-0.0012584915174 -9.4789004189858i$\\
        $1$& $-0.0266639257692i -0.0235576485732$ & $0.0012417895565 +0.0537580805588i $\\
       $2$&  $ -1.62952996537[4]i+4.09974571511[4]$ &$1.77927321993[5] + 3.62538594870[4]i $\\
         \hline 
           $n$ & 
        $\Pi_{q,1}^{(n)}$&$\hat{\Pi}_{q}^{(n)}$\\
         \hline
         $0$& $4.3446402030184+ 15.4257466285002 i $&$ - 0.0012585520475i +9.4789002636561$\\
        $1$& $0.0266639255426 - 0.0235576487255 i $ & $-0.0012417898853  i+0.0537580793133$\\
       $2$&  $-1.62952995392[4]-4.09974572875[4]i $ &$1.77927341291[5]i-3.62538588259[4]$\\
         \hline
         \end{tabular}
        \caption{The expansions of the quantum periods with $F=0.03$, $m=\frac{1}{2}$ for $\xi$-equation.}
     \label{table:singularqp3}
\end{table}

We find that the situation is simplified for a special choice of $\ell$, where the quantum periods follow compact TBA equations for the double-well potential. Firstly, one notes that under the coordinate transformation $x\to z^2$ and rescaling of the function $\psi(x)\to z^{\frac{1}{2}}\phi(z)$, the original equation \eqref{eq:sseq} becomes
\begin{equation}
\label{eq: quartic}
   \left(-\hbar^2\frac{d^2}{dz^2}+\left(4u_0 z^4+4u_1z ^2+4u_2\right)+\hbar^2\frac{16\ell(\ell+1)+3}{4z^2}\right) \phi(z)=0.
\end{equation}
For $\ell=-\frac{1}{4},-\frac{3}{4}$, or equivalently $m=\frac{1}{2}$, the centrifugal term vanishes. The original equation is mapped to one with a double-well potential, whose TBA equations are already established, for example in \cite{IMS, Emery}. We denote the quantum periods related to \eqref{eq:sseq} as $\Pi_{s,1}^{(n)}$ for one-cycle $\gamma$ and $\hat{\Pi}_{s}^{(n)}$ for one-cycle $\hat{\gamma}_1$ defined at the beginning of this section. Quantum periods related to \eqref{eq: quartic} are denoted as $\Pi_{q,1}^{(n)}$ and $\hat{\Pi}_{q}^{(n)}$ for corresponding cycles. 

We consider a specific case $F=0.03$ with $m=\frac{1}{2}$. The turning points for \eqref{eq: quartic} are $\pm \sqrt{e_1}$ and $\pm \sqrt{e_2}$ with $e_{1,2}$ the turning points for \eqref{eq:sseq}. Then the classical periods are defined as
\begin{equation}
  \begin{aligned}
      \Pi_{q,1}^{(0)}&=2\int_{-\sqrt{e_1}}^{-\sqrt{e_2}}\sqrt{Q_0^q(x)} dx,\quad \hat{\Pi}_{q}^{(0)}=2\int_{-\sqrt{e_1}}^{\sqrt{e_1}}\sqrt{Q_0^q(x)} dx,\, \text{for} \,\xi \,\text{variable},\\
       \Pi_{q,1}^{(0)}&=2\int_{-\sqrt{e_2}}^{-\sqrt{e_1}}\sqrt{Q_0^q(x)} dx,\quad \hat{\Pi}_{q}^{(0)}=2\int_{\sqrt{e_1}}^{-\sqrt{e_1}}\sqrt{Q_0^q(x)} dx,\, \text{for} \,\eta \,\text{variable},
  \end{aligned}
\end{equation}
here $Q_0^q(x)=4u_0 z^4+4u_1z ^2+4u_2$ is the quartic potential.
The higher orders can be evaluated from the Picard-Fuchs method as well. let us now consider the TBA equations for \eqref{eq: quartic}. For parameters in $\eta$ variable, the corresponding TBA equations are in the minimal chamber as
\begin{equation}
\label{eq:minitba}
\begin{aligned}
& \log Y_1(\theta)=-m_1 \mathrm{e}^\theta+\int_{\mathbb{R}} \frac{\log \left(1+\hat{Y}\left(\theta^{\prime}\right)\right)}{\cosh \left(\theta-\theta^{\prime}\right)} \frac{\mathrm{d} \theta^{\prime}}{2 \pi}, \\
& \log \hat{Y}(\theta)=-\hat{m} \mathrm{e}^\theta+2 \int_{\mathbb{R}} \frac{\log \left(1+Y_1\left(\theta^{\prime}\right)\right)}{\cosh \left(\theta-\theta^{\prime}\right)} \frac{\mathrm{d} \theta^{\prime}}{2 \pi} .
\end{aligned}
\end{equation}
\begin{table}[h]
        \centering
        \begin{tabular}{c c c}
         \hline 
         $n$ & $m_1^{(n)}$ & $\hat{m}^{(n)}$\\
          \hline 
         $0$ & $0.4640679572901 - 0.6555218319259 i$&  $ - 0.4550034327887 -8.8968672576106i$\\
        $1$ &  $0.0007954104344 - 0.054482181983i$ &    $0.3863907823625 + 0.5839673005070i$ \\
         $2$ & $ 0.0001177964828-0.0015026546306 i$  & $-2.2800851644720 + 0.6221615219336 i $\\
          \hline
        $n$&   $\Pi_{s,1}^{(n)}$&$2\hat{\Pi}_{s}^{(n)}$\\
         \hline
         $0$& $0.4640679650555  i+0.6555217872253  $&$-0.4550034419775 - 8.8968672279807 i$\\
        $1$& $ -0.0007954088972 i-0.0544821836255$ & $-0.3863907817713 - 0.5839672972422i$\\
       $2$&  $0.0001178053284 i+0.0015026585094$ &$-2.2800851547326 + 0.6221615231299 i$\\
        \hline 
           $n$ & 
        $\Pi_{q,1}^{(n)}$&$\hat{\Pi}_{q}^{(n)}$\\
         \hline
         $0$& $0.4640679052528 - 0.6555218495485i$&$0.4550034203263i-8.8968672224660  $\\
        $1$& $0.0007954117778 - 0.0544821822247i$ & $-0.3863907794426i+0.5839672977885  $\\
       $2$&  $0.0001177955168 - 0.0015026454486i$ &$2.2800851542317i +0.6221615160777 $\\
         \hline
         \end{tabular}
        \caption{The expansions of the quantum periods with $F=0.03$, $m=\frac{1}{2}$ for $\eta$-equation.}
     \label{table:singularqp4}
\end{table}
The TBA equations for the $\xi$ variable are now in the maximal chamber. Equation (5.21) in \cite{IMS} gives the explicit form of these equations, the mass parameters are related by
$m_2=\hat{m}, m_{12}=m_1-i\hat{m}$, and $m_{123}=2m_1-i \hat{m}$ while $m_1$ is the same as we used in this paper. Then we found the quantum periods and the large $\theta$ expansion of the quartic TBA equations are related as follows
\begin{equation}
\label{eq:quartictba}
\ba
    &\Pi_{q,1}^{(n)}= (-1)^{n+1}i \Pi_{s,1}^{(n)}= m_1^{(n)},\quad (-1)^{n+1}i \hat{\Pi}_{q}^{(n)}=2\hat{\Pi}_{s}^{(n)}=\hat{m}^{(n)},\,\text{for}\, \xi \text{-equation},\\
    &\Pi_{q,1}^{(n)}=(-1)^{n+1}i\Pi_{s,1}^{(n)}= m_1^{(n)},\quad (-1)^{n}i\hat{\Pi}_{q}^{(n)}= 2\hat{\Pi}_{s}^{(n)}=(-1)^{n}\hat{m}^{(n)},\,\text{for}\, \eta\text{-equation}.
\ea
\end{equation}
Table \ref{table:singularqp3} shows the numerical calculations for $\xi$-equation with $F=0.03$ and $m=\frac{1}{2}$, where corresponding TBA equations are in the maximal chamber, and table  \ref{table:singularqp4} gives results for $\eta$-equation, with its TBA equations \eqref{eq:minitba} in the minimal chamber. Numerical calculations confirm the correspondence \eqref{eq:quartictba} very well. 

In this section, we consider TBA equations related to the Stark problem without Langer's modification. These equations on the one hand confirm the validity of the ODE/IM correspondence and give an alternative computational approach for quantum periods on the other hand. Unfortunately, we have not found consistent EQCs in terms of these quantum periods without Langer's modification. 

\section{Discussion and conclusion}
\label{sc:concl}

In this paper, we have applied the exact WKB analysis to the well-known Stark problem to resolve the energy frequency and ionization rate simultaneously. This method has several advantages. 
\begin{itemize}
    \item First, it gives the complex resonant frequency including $E_c$ and ionization rate at the same time, and it applies to a wide range of parameter choices for $F$ and $m$.
    \item Second, the analysis is exact as the term EWKB implies. We have derived the exact quantization conditions by utilizing Stokes graphs and found the BS quantization condition is already exact in the specific parameter regions, which clarifies the validity of the WKB approximation and quantization conditions in \cite{starkwkb69,starkwkb82,starkwkb99}. 
    \item Third, the method presented in this paper is fully analytic in principle, except for the final numerical calculation, which makes it possible to investigate the asymptotic behaviour in the limit of the parameters. 
    \item Finally, this method can be systematically extended to higher orders, which is crucial to knowing the large-order behaviour of the asymptotic expansions of the quantum periods even its non-perturbative contributions.
\end{itemize}
 We have examined our formulation by numerical calculations which are compared with the results by the Riccati-Padé method in \cite{rpm} to find agreements. It is proved that this method including Langer's modification and quantum corrections gives a complete resolution of the Stark problem.

In addition, we have discussed the TBA systems related to the Stark problem in the absence of Langer's modification. It gives much more information on the quantum periods by the ODE/IM correspondence. We confirmed the corrections introduced in \cite{xiyin} by comparing the expansions of the logarithm of the $Y$-functions with quantum periods computed from the Picard-Fuchs method. We also discussed the wall-crossing of the TBA equations during the parameters change and wrote down the corresponding quartic TBA equations for special $\ell$.
All these equations are tested by numerical calculations. However, there are still some aspects to be improved. 
\begin{itemize}
    \item First, we have not found self-consistent EQCs for all parameters for the bare equation in section \ref{sc:tba} without Langer's modification from the corresponding Stokes graphs, which motivated us to consider the Stark problem at the beginning. We expect to achieve this to a large extent since this method is exact if all-order quantum corrections are included, and TBA equations are applicable in this prescription. The failure might come from the non-matching of wave functions at the semi-classical level, which shows the essence of Langer's modification in the WKB method. 
    \item Second, it seems not easy to directly get the explicit formula for energy frequency from our EQCs to compare with other perturbative formulas. We are also expecting to find a compact form of the TBA equations after wall-crossing as in the case of polynomial potentials. There are some references on this topic like \cite{cluster1,cluster2} or from the GMN formalism \cite{GMN}, see \cite{ouyang} for example. TBA equation with both $\ell$ correction and wall-crossing included is also an open question. It is possible to write down the TBA equations including Langer's modification, but their complete forms are still lacking. We would like to consider them in our future work.
\end{itemize}
The present WKB and TBA approach can also be applied to a variety of fields such as the quasi-normal modes of black holes and the quantum Seiberg-Witten curve, which opens up the potential of our method  \cite{Fioravanti, Aminov:2020yma, Ito:2019twh}.

\section*{Acknowledgements}

We would like to thank Keita Imaizumi for his valuable discussions at the early stage. We also thank Hongfei Shu for his kind correspondence. The work of J.Y. is partly supported by the Advanced Research Center for Quantum Physics and Nanoscience, Tokyo Institute of Technology.
The work of K.I. is supported in part by Grant-in-Aid for Scientific Research 21K03570 from Japan Society for the Promotion of Science (JSPS).


\end{document}